\documentclass[twocolumn]{aastex62}
\usepackage{enumitem}
\usepackage{graphicx,url}
\usepackage{color}
\usepackage{amsmath}
\usepackage{mathtools}

\def\j1550{XTE~J1550$-$564}

\def\rxte{{\it RXTE}}

\def\nustar{{\it NuSTAR}}

\def\asca{{\it ASCA}}

\submitjournal{ApJ}

\shorttitle{Reflection Spectroscopy of \j1550}
\shortauthors{Connors et al.}

\setlist[itemize]{leftmargin=*}

\begin{document}

\title{Conflicting disk inclination estimates for the black hole X-ray binary \j1550}

\correspondingauthor{Riley~M.~T.~Connors}
\email{rconnors@caltech.edu}

\author{Riley~M.~T.~Connors}
\affiliation{Cahill Center for Astronomy and Astrophysics, California Institute of Technology, \\
  Pasadena, CA 91125, USA}

\author{Javier~A.~Garcia}
\affiliation{Cahill Center for Astronomy and Astrophysics, California Institute of Technology, \\
  Pasadena, CA 91125, USA}
  \affiliation{Dr Karl Remeis-Observatory and Erlangen Centre for Astroparticle Physics,\\
  Sternwartstr. 7, D-96049 Bamberg, Germany}
    
\author{James~F.~Steiner}
\affiliation{MIT Kavli Institute, 77 Massachusetts Avenue, 37-241, \\
Cambridge, MA 02139, USA}
\affiliation{CfA, 60 Garden St. Cambridge, MA 02138, USA}

\author{Victoria Grinberg}
\affiliation{Astronomy and High Energy Astrophysics, Institute for Astronomy und Astrophysics, \\
 Tuebingen, 72076, Germany}
 
\author{Thomas Dauser}
\affiliation{Dr Karl Remeis-Observatory and Erlangen Centre for Astroparticle Physics,\\
  Sternwartstr. 7, D-96049 Bamberg, Germany}
  
\author{Navin Sridhar}
\affiliation{Department of Astronomy, Columbia University,\\
    550 W 120th St, New York, NY 10027, USA}
    
\author{Efrain Gatuzz}
\affiliation{ESO,Karl-Schwarzschild-Strasse 2, D-85748 Garching bei München, Germany}

\author{John Tomsick}
\affiliation{Space Sciences Laboratory, University of California Berkeley,\\
7 Gauss Way, Berkeley, CA 94720-7450}

\author{Sera B. Markoff} 
\affiliation{Anton Pannekoek Institute for Astronomy, University of Amsterdam,\\
 Science Park 904, 1098 XH Amsterdam, The Netherlands}
 
 \author{Fiona Harrison}
 \affiliation{Cahill Center for Astronomy and Astrophysics, California Institute of Technology, \\
  Pasadena, CA 91125, USA}








\begin{abstract}
\j1550 is a black hole X-ray binary for which the dynamical characteristics are well established, and the broadband spectral evolution of the source has been well studied. Its orbital inclination is known to be high, at $\sim75^{\circ}$, with the jet estimated to align well with the orbital axis. We explore simultaneous observations made with \asca\ and \rxte, covering the $1$--$200$~keV band, during the early stages of the first outburst of \j1550\ in its hard-intermediate state, on 1998-09-23/24. We show that the most up-to-date reflection models, applied to these data, yield an inclination estimate much lower than found in previous studies, at $\sim40^{\circ}$, grossly disagreeing with the dynamically estimated orbital inclination. We discuss the possible explanations for this disagreement and its implications for reflection models, including possible physical scenarios in which either the inner disk is misaligned both with binary orbit and the outer jet, or either the inner accretion flow, corona, and/or jet have vertical structure which leads to lower inferred disk inclination through various physical means.

\end{abstract}


%
%
%
\section{Introduction}\label{sec:intro}
The relative inclinations of the spin/jet/disk axes and orbital plane of X-ray binaries with respect to the line of sight are much sought after quantities (see, e.g., \citealt{Hjellming1995,Orosz1997,Orosz2001,Fragos2010}). Such estimates not only inform our understanding of how X-ray binaries form and evolve, they are also imperative for the modeling of X-ray reflection off accretion disks around compact objects, a common tool for estimating key accretion properties, as well as black hole spin (see, e.g., \citealt{Ross2005,Ross2007,FabianRoss2010,Dauser2014,Garcia2014,Fabian2014}), and more pertinently the relativistically-distorted thermal disk continuum \citep{Zhang1997}. The modeling of jet emission also relies upon inclination measurements, since beaming depends inherently on the jet orientation. The black hole X-ray binary (BHB) \j1550\ is an example of a system in which the independent estimates of its jet and orbital inclinations agree well within the uncertainties, making it a seemingly good test case for reflection studies. 

\j1550 was discovered by the All-Sky Monitor on board the \textit{Rossi X-ray Timing Explorer} (\rxte) on September 6 1998 \citep{Smith1998} as a transient Galactic X-ray source. Pointed daily observations with \rxte\ subsequently monitored its outburst over an eight-month period \citep{Sobczak2000}, two weeks into which it exhibited a dramatic increase in X-ray flux up to 7 Crab. Following this X-ray flare, radio observations with the Australian Long Baseline Array (ALBA) revealed a large-scale superluminal jet propagating both east-and-westward from the X-ray source \citep{Hannikainen2009}. Two years later the jet was observed in X-rays and shown to be decelerating \citep{Corbel2002b,Kaaret2003,Tomsick2003}. More recent observations of the jet of \j1550\ have revealed that its morphology is evolving, with the X-ray jet continuing to expand \citep{Migliori2017}. In addition to the characterization of its radio properties, optical/IR observations of \j1550\ have yielded a reliable dynamical model for the system, with BH mass, distance, orbital period, and inclination estimates of $M_\mathrm{BH}=9.1 \pm 0.6~M_\odot$, $D=4.4^{+0.6}_{-0.4}~\mathrm{kpc}$, $P_\mathrm{orb}=1.54~\mathrm{days}$ and $i=75^{\circ} \pm 4^{\circ}$ respectively \citep{Orosz2002, Orosz2011}. During the 1998/99 outburst, quasi-periodic oscillations (QPOs) are clearly detected in multiple \rxte\ observations of the source \citep{Remillard2002a}. Since the first detected outburst in 1998, \j1550\ has gone into outburst four more times, and has been observed in the plethora of X-ray spectral and timing states (see, e.g., \citealt{rm06} for a review of X-ray binary states). \j1550\ underwent a shorter yet complete outburst in 2000 \citep{Rodriguez2003}, and three subsequent ``failed" outbursts in 2001, 2002, and 2003, in which the source did not transition into the soft state \citep{rm06}. Thus, the X-ray spectral and timing characteristics and evolution of the source have been explored in great detail \citep{Sobczak2000,Homan2001,Remillard2002b, Kubota2004,Dunn2010}. 

X-ray spectral modeling of \j1550\ with the goal of estimating the spin of the black hole (BH) has been conducted using several methods \citep{Davis2006,Miller2009b,Steiner2011}: direct modeling of the thermal disk continuum with the relativistic thin accretion disk models \texttt{kerrbb} and \texttt{kerrbb2} \citep{Li2005,McClintock2006}, modeling of the Fe K emission line with the reflection model \texttt{refbhb} (\citealt{Ross2007}; a variant on the reflection model \texttt{reflionx}, \citealt{Ross2005}, which includes a thermal disk continuum as an irradiative component), relativistically smeared using the kernel \texttt{kerrconv} \citep{Brenneman2006}, and modeling of QPOs \citep{Motta2014}. The spin constraints are all low, leading to a rough constraint of $a_\mathrm{\star}=0.5$. 

All these methods used to determine black hole spin rely on an estimate of the orbital inclination that is accurate, as well as the assumption that this inclination represents the inclination of the inner edge of the accretion disk with respect to the spin axis. \cite{Steiner2011} applied both continuum and reflection methods whilst making use of the most up-to-date dynamical measurements of the binary system, and they found spin measurements which agree under the assumption that the disk inclination is within $1\sigma$ of the estimated $i=75^{\circ}$. \cite{Steiner2012} subsequently showed, through modeling of the large-scale jet observed during the 1998 outburst of \j1550,  that the jet and orbital inclination are aligned to within $12^{\circ}$ at 90\% confidence. 

Despite the exhaustive nature of the modeling conducted by \cite{Steiner2011}, there exists still a lack of modeling of strictly simultaneous broadband X-ray spectral observations of \j1550. The thermal continuum modeling in their work was applied to \rxte-PCA (Proportional Counter Array) data only, and their modeling of the broadened Fe K line was conducted on observations taken with the \textit{Advanced Satellite for Cosmology and Astrophysics} (\asca), with both datasets overlapping during September 12 and 23--24 1998. Other attempts to model these simultaneous data have discussed the difficulty in modeling the simultaneous 25~ks \asca\ and 3~ks \rxte\ observation on September 23--24. Some instead chose to optimize photon statistics and model quasi-simultaneous observations consisting of combined PCA spectra covering the September 23--October 6 period that followed \citep{Gierlinski2003, Hjalmarsdotter2016}. 

We note that whilst the latter works focused on reproducing broadband X-ray spectral features related to pair-production, the fact remains that no previous studies of the broadband spectrum of \j1550\ has been able to explain the simultaneous \asca\ and \rxte\ data available. In this paper we present reflection modeling of the strictly simultaneous \asca\ and \rxte\ (both PCA and HEXTE---High Energy X-ray Timing Experiment) observations during September 12 and 23--24 1998, and show that there is an offset in the slope of the spectra of the \asca-GIS and \rxte-PCA instruments of the order $\sim0.1$ in photon index. We show that this cross-calibration difference can either be accounted for via a shift in the channel-to-energy gain of the \asca\ GIS (Gas Imaging Spectrometer) 2 and GIS 3 detectors, or by introducing a cross-calibration slope offset between the GIS and PCA spectra. We subsequently show that after accounting for these differences, reflection modeling of these simultaneous data lead to an inclination constraint much lower than those adopted in the continuum and reflection modeling of \cite{Steiner2011} and those derived from dynamical modeling of orbit and jet of \j1550\ \citep{Orosz2011,Steiner2012}. 

This paper is structured as follows: in Section~\ref{sec:data} we introduce the \asca\ and \rxte\ data and how it is reduced prior to modeling, and in Section~\ref{sec:imod} we show results of spectral modeling of the individual datasets. In Section~\ref{sec:smod} we show results of simultaneous spectral modeling of the \asca\ and \rxte\ data, arriving at a final model of relativistic reflection. In Section~\ref{sec:discussion} we discuss the implications of our modeling, focusing in particular on the details of the low disk inclination we measure. Finally in Section~\ref{sec:conclusion} we draw our conclusions. 

\begin{figure}[!ht]
\includegraphics[width=\linewidth]{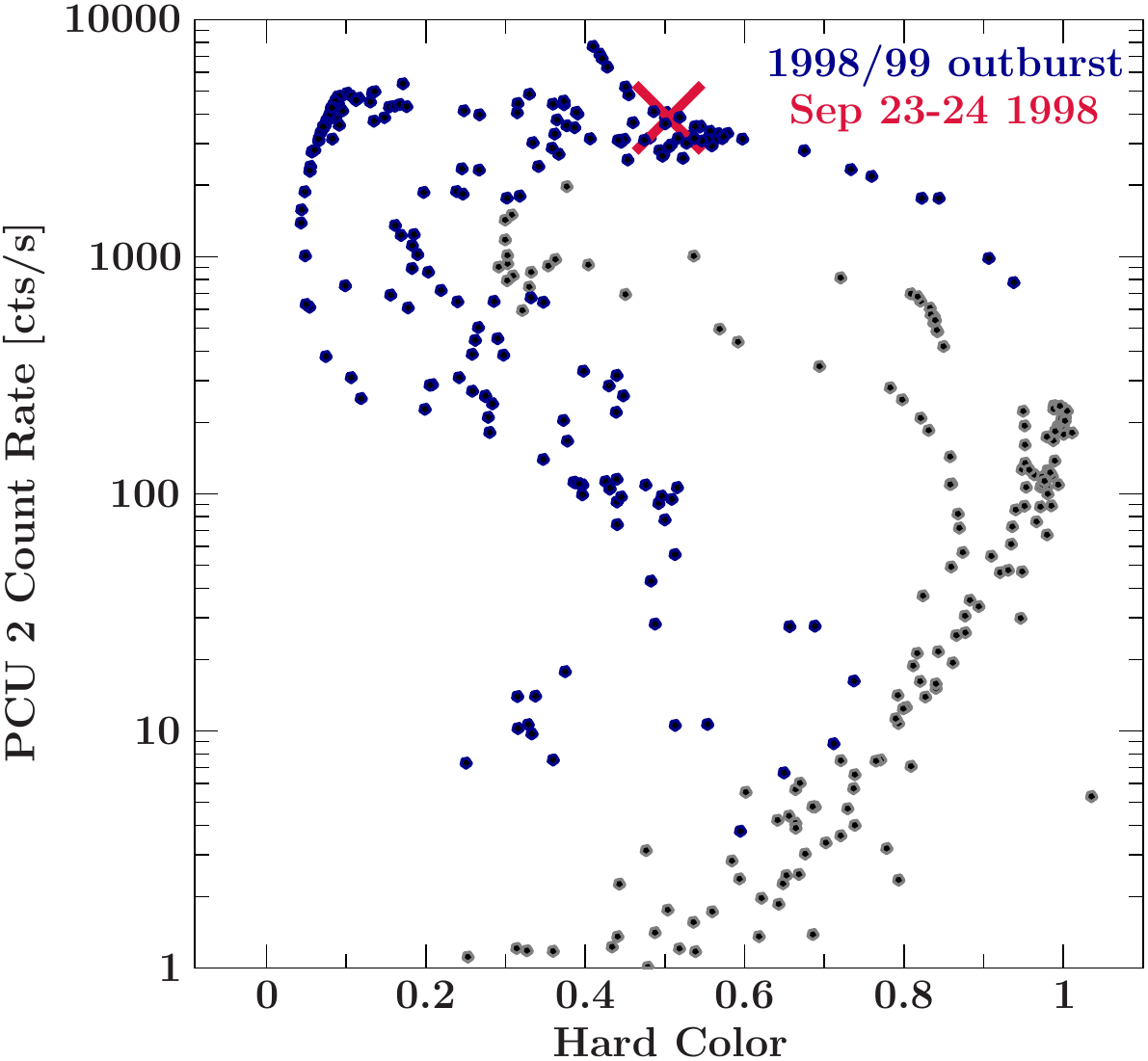}
\caption{Hardness-intensity diagram including all \rxte\ observations of \j1550. The hard color is defined as the ratio of source counts in the hard and soft bands, $[8.6$--$18]/[5$--$8.6$]~keV. The first outburst detected by \rxte\ is shown in dark blue. The observation window focused on in this work is shown in crimson, a 25~ks \asca\ exposure from the night of Sep 23 into Sep 24, along with a 3~ks \rxte\ exposure.}
\label{fig:hid}
\end{figure}

\begin{figure}[!ht]
\includegraphics[width=\linewidth]{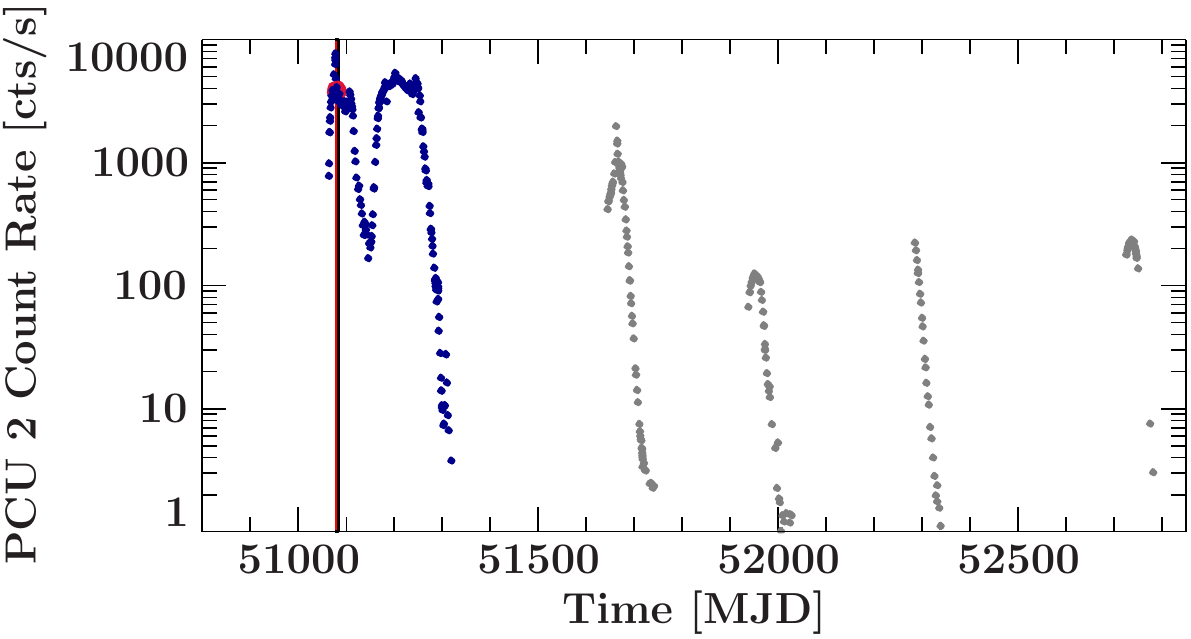}
\includegraphics[width=\linewidth]{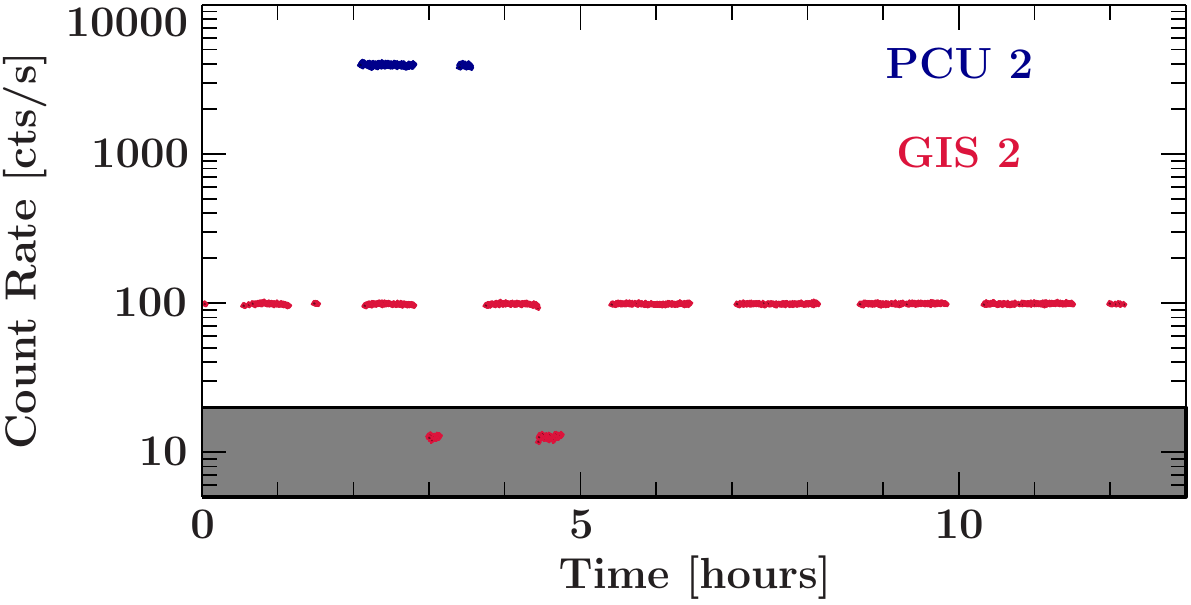}
\caption{{\bf Top:} Long term \rxte\ light curve of \j1550 highlighting the 1998/99 outburst in blue, with a crimson line indicating the Sep 23--24 observation simultaneous with \asca\ observations of the source. {\bf Bottom:} The expanded $\sim13$-hour simultaneous \rxte-PCA and \asca-GIS light curves on Sep 23--24 1998. The gray region highlights excluded dips in GIS counts due to occultation.}
\label{fig:lcs}
\end{figure}

%
%
%
 \section{Data reduction}
 \label{sec:data}
\asca\ observed \j1550 twice shortly after it was discovered as a transient outbursting source in September 1998, on September 12 and 23--24. It was also observed again in March 1999. During all three of these \asca\ observations there was simultaneous \rxte\ coverage. We focus here on the September 23--24 observation when \j1550\ was in a hard-intermediate state, ObsID 15606010 (\asca), with an exposure of 25~ks and a total of $4.4\times10^6$ counts, such that we can compare with previous works on these particular observations (e.g., \citealt{Kubota2004,Steiner2011,Hjalmarsdotter2016}). The simultaneous \rxte\ coverage consists of a $\sim3$~ks PCA exposure with $10^7$ counts, and a $\sim1$~ks HEXTE exposure with $1.4\times10^5$ cluster A counts, and $9.8\times10^4$ cluster B counts, ObsID 30191-01-10-00. The position of these observations in the X-ray hardness-intensity space is shown in Figure~\ref{fig:hid}. Next, we describe how we reduce and model the \asca\ and \rxte\ data, and how this compares with previous attempts to characterize these data and model the broadband spectrum. 

\subsection{\asca}
Using the tool \texttt{xselect}, \texttt{heasoft-v6.22.1}, we extracted GIS 2 and GIS 3 spectra from the available archival screened (using standard screening) \asca\ event files. We first extracted a source spectrum from a circular region of 6-arcmin radius, centered on the source. We then extracted an off-source spectrum of equal size to represent the \asca\ background. Upon inspection of both the GIS 2 and GIS 3 light curves, we noticed dips in the count rate which appeared to be left after applying standard filtering (see Figure~\ref{fig:lcs}), likely explained by Earth occultation during satellite orbit. We manually excluded these portions of the light curve. We obtained the response matrix files for GIS 2 and GIS 3 respectively, \texttt{gis2v4\_0.rmf} and \texttt{gis3v4\_0.rmf}, from the archive and used the FTOOL \texttt{ascaarf} to generate an ancillary response file for the selected region of the source spectrum. Finally, we corrected the GIS spectra for dead time using the FTOOL \texttt{deadtime} in accordance with \cite{Makishima1996}. During extraction we grouped the source spectrum by a factor of 4 such that there are 256 channels in both the GIS spectra. The GIS 2 and GIS 3 spectra were then further combined into an average GIS spectrum, ancillary response and background using the tool \texttt{addascaspec}. This combination is motivated by the fact that the GIS calibration against Crab spectra was performed using a coadded GIS 2 + GIS 3 spectrum (Yoshihiro Ueda, private communication). 




\begin{figure*}[ht!]
\centering
\hspace{0.5cm}
\includegraphics[width=0.8\linewidth]{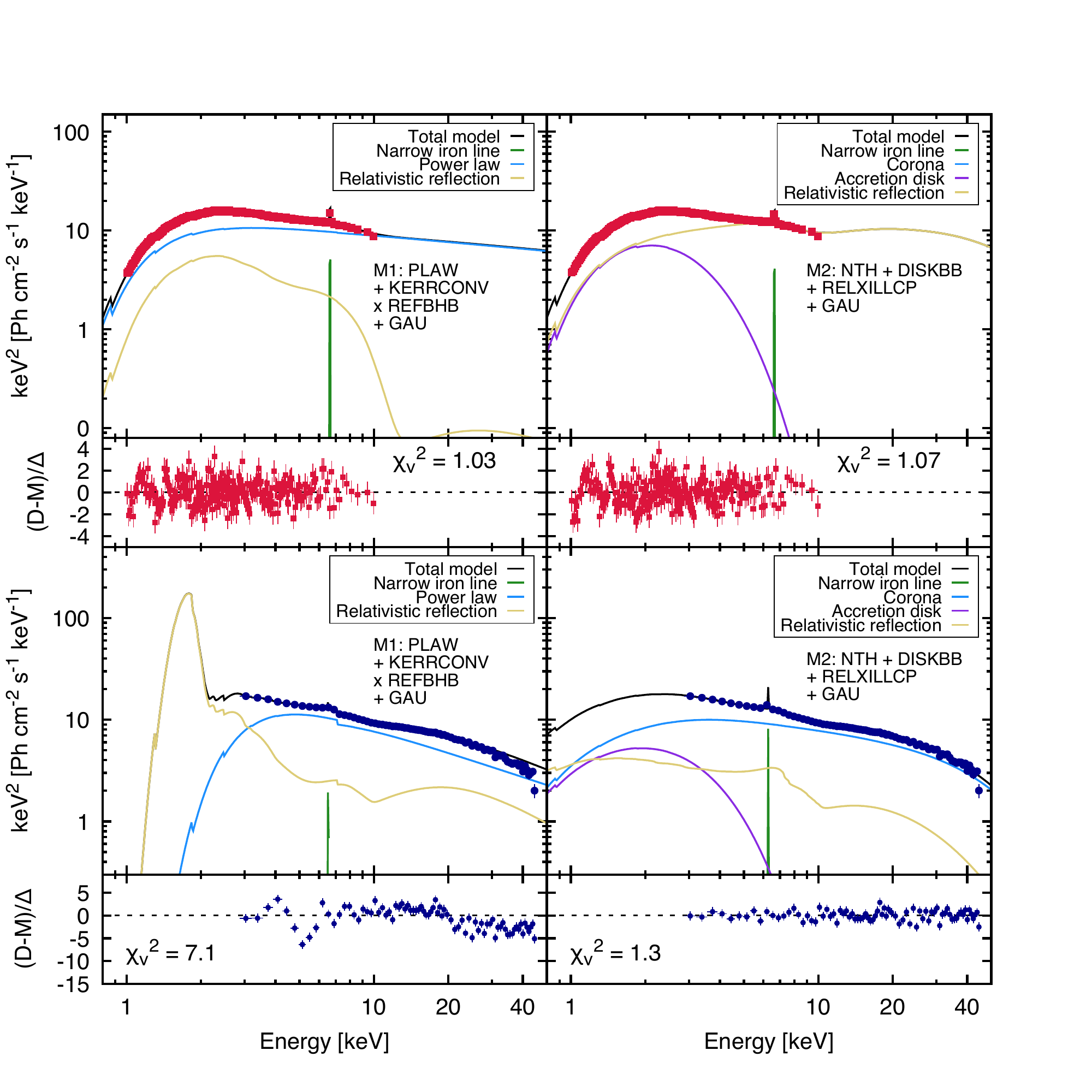}
\caption{Individual fits to ASCA and PCA observation of \j1550 on September 23--24 1998. The spectra from GIS 2 and GIS 3 are combined into an averaged GIS spectrum, shown in crimson, and the PCA spectrum is shown in dark blue.The key shows the colors of the model components. Two model classes are considered, 1 and 2, comparing the reflection models \texttt{refbhb} and \texttt{relxillCp}. The lower panels of each fit display the data(D)$-$model(M) residuals, normalised by the data uncertainties ($\Delta$).}
\label{fig:indfits}
\end{figure*}


\begin{figure}[ht!]
\includegraphics[width=\linewidth]{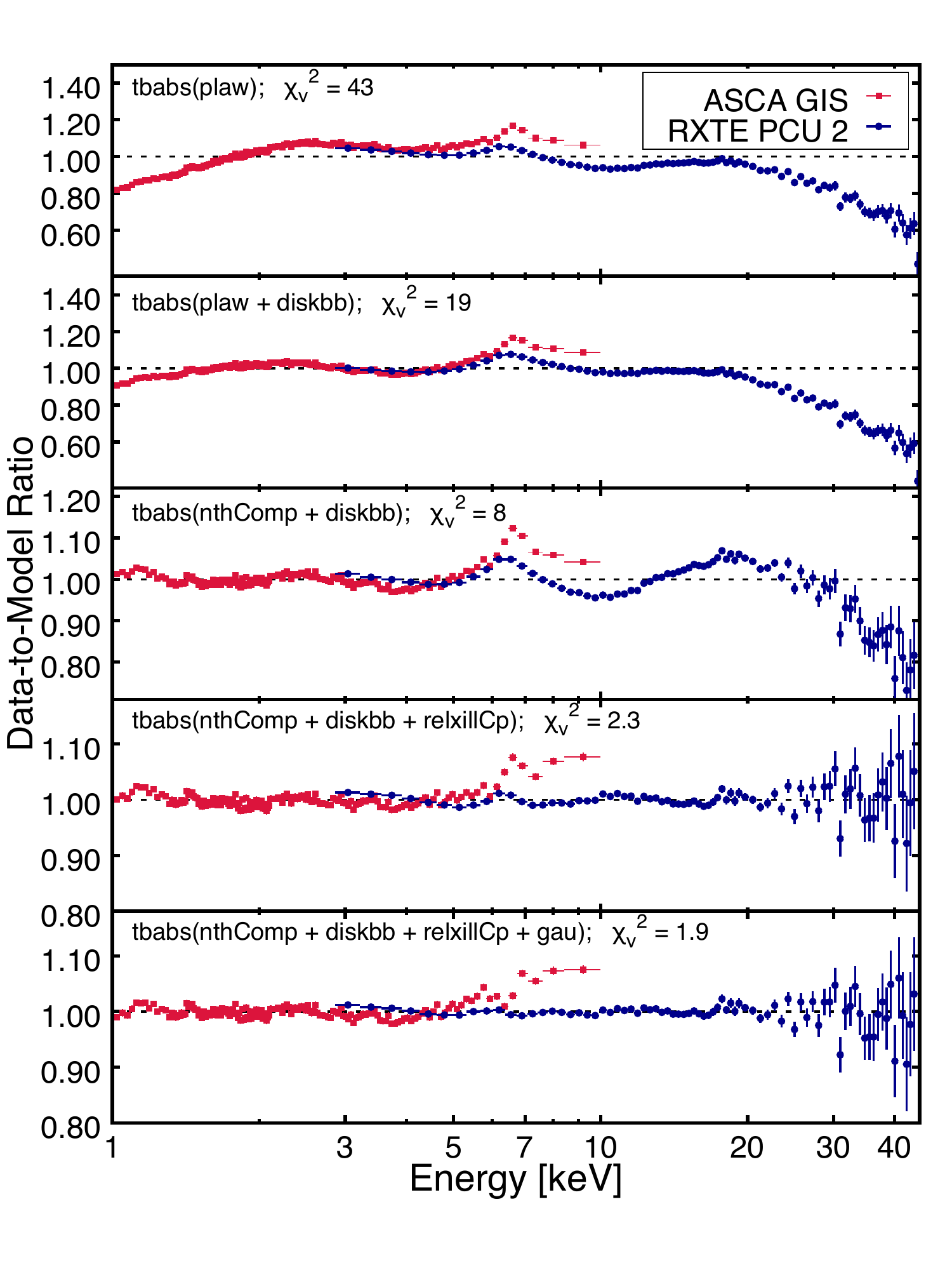}
\caption{Progression of model fits to the simultaneous \asca-GIS (red squares) and \rxte-PCA (blue points) observations of \j1550\ on September 23--24 1998. One can see that an offset in the slope persists at every stage of the model progression. Residuals are shown as a ratio between the data and model counts.}
\label{fig:gis_pca_offset}
\end{figure}

\begin{figure*}[ht!]
\includegraphics[width=0.5\linewidth]{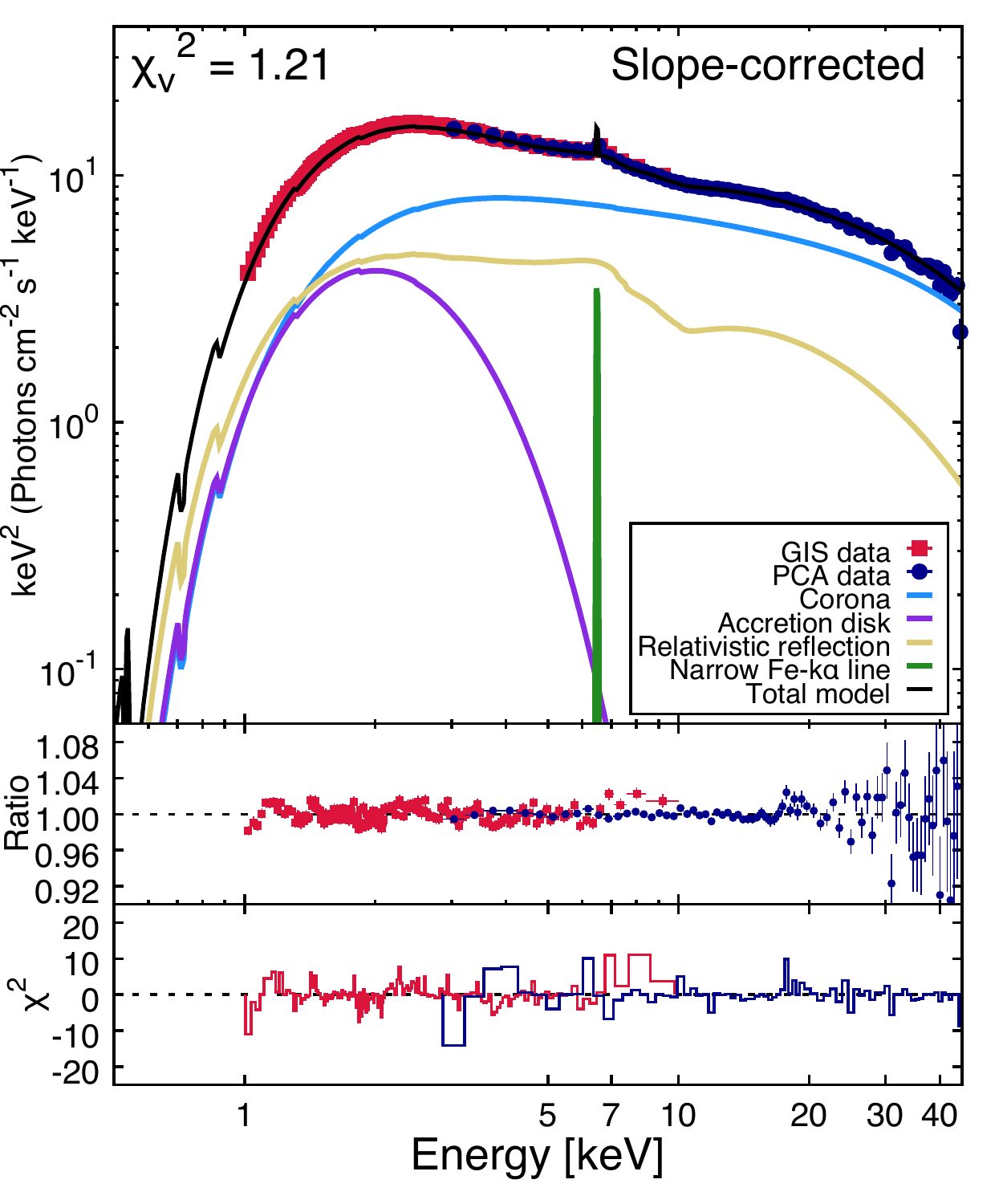}
\includegraphics[width=0.5\linewidth]{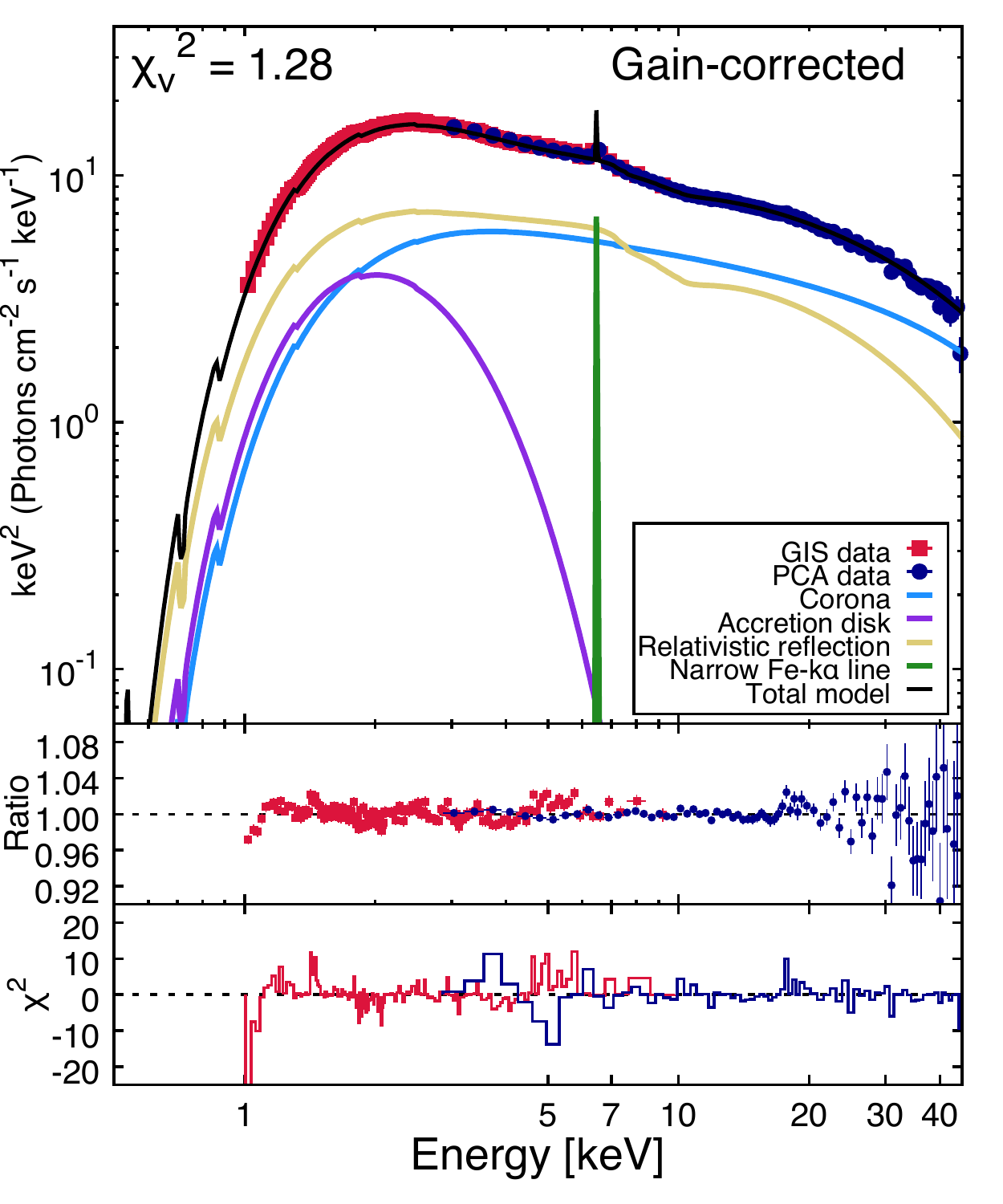}
\caption{Fits of Model~2 to the simultaneous \asca-GIS and \rxte-PCA spectra taken on September 23--24 1998, showing the effect of applying a) a slope correction to the PCA data with respect to the Crab spectrum, using the \texttt{crabcorr} model, and b) a gain shift correction to the response of the GIS detector. The individual model components are shown along with the unfolded spectra, and the panels below show the data-to-model ratio and $\chi^2$ residuals respectively. }
\label{fig:gis_pca_corrected}
\end{figure*}

\begin{figure}[ht!]
\includegraphics[width=\linewidth]{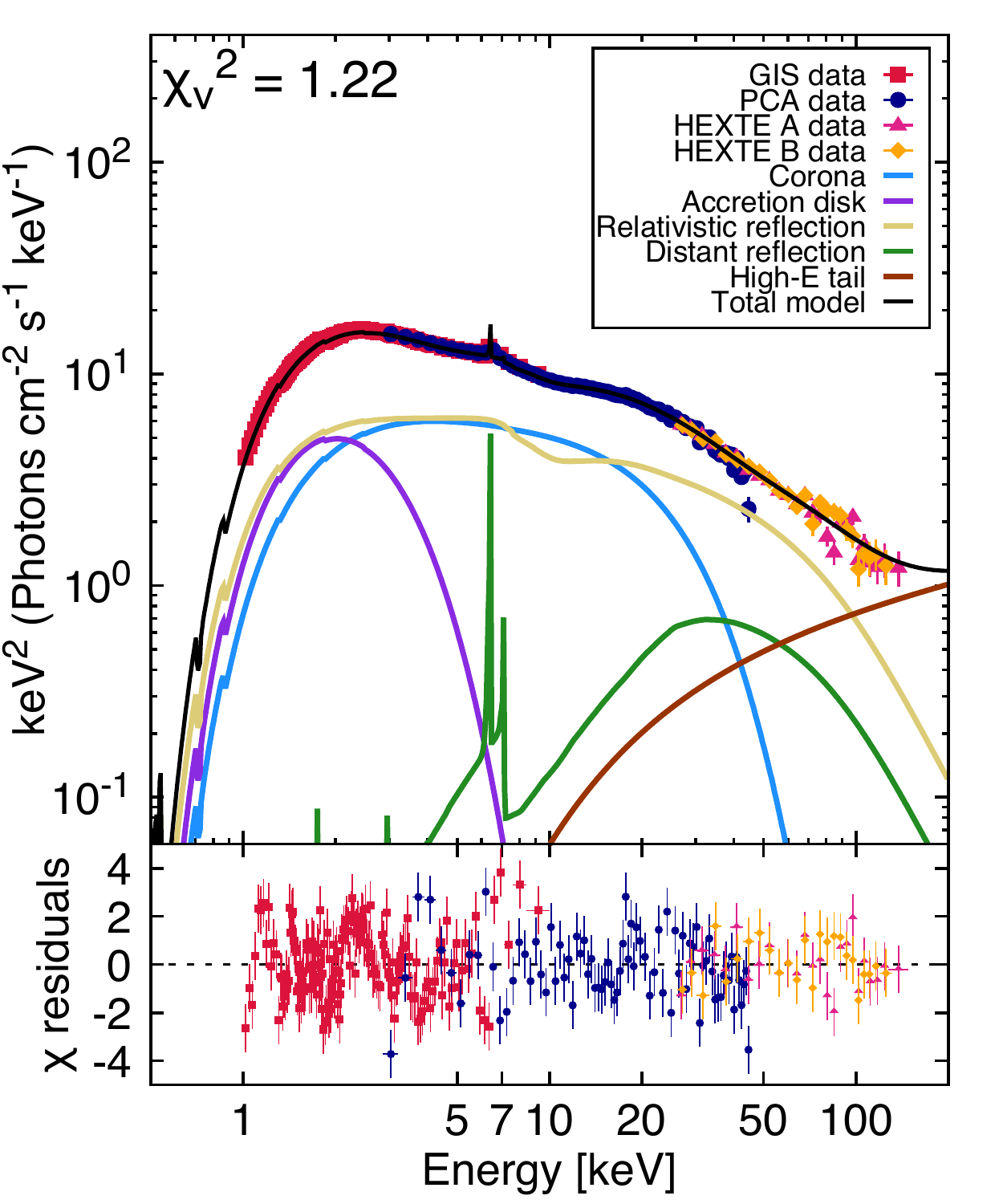}
\caption{Fits of Model~3 to the simultaneous \asca-GIS, \rxte-PCA-and-HEXTE spectra taken on September 23--24 1998. The individual model components are shown along with the unfolded spectra, and the panels below show the data$-$model residuals, normalized by the data uncertainties. }
\label{fig:finalfit}
\end{figure}

\def\mIAcnorm{$0.97^a$}			
\def\mIAcdelgam{$-0.01^a$}			
\def\mIAnhI{$0.64^{+0.02}_{-0.02}$}		
\def\mIAga{$2.22^{+0.04}_{-0.08}$}			
\def\mIApln{$14 \pm 2$}			
\def\mIAq{$2.5^{+0.3}_{-0.1}$}			
\def\mIAa{$0.6^{+0.2}_{-0.2}$}			
\def\mIAi{$84 \pm 3$}			
\def\mIAHden{$1.00_{-0.04}$}			
\def\mIAbbkT{$0.544^{+0.002}_{-0.007}$}			
\def\mIAillum{$0.24^{+0.13}_{-0.05}$}			
\def\mIAren{$0.37^{+0.17}_{-0.10}$}			
\def\mIAC{$0.984^{+0.002}_{-0.002}$}			
\def\mIAgn{$0.011^{+0.001}_{-0.001}$}			
\def\mIAge{$6.64^a$}			
\def\mIAchi{$776$}			
\def\mIAdof{$752$}			
\def\mIAredc{$1.03$}			

\def\mIIAcnorm{$0.97^a$}			
\def\mIIAcdelgam{$-0.01^a$}			
\def\mIIAnhI{$0.605^{+0.009}_{-0.021}$}		
\def\mIIAga{$1.84^{+0.05}_{-0.07}$}			
\def\mIIAkTe{unconstrained}			
\def\mIIAnthn{$<1$}			
\def\mIIAkTbb{$0.673^{+0.016}_{-0.008}$}			
\def\mIIAdin{$6700^{+300}_{-500}$}			
\def\mIIAa{$0.5^a$}			
\def\mIIAq{$3^a$}			
\def\mIIAi{$23^{+8}_{-20}$}			
\def\mIIAreAfe{$10_{-3}$}			
\def\mIIArexi{$4.5^{+0.1}_{-0.1}$}			
\def\mIIAren{$0.038^{+0.002}_{-0.007}$}			
\def\mIIAgn{$0.008^{+0.002}_{-0.002}$}			
\def\mIIAge{$6.20^{+0.13}_{-0.06}$}			
\def\mIIAchi{$819$}			
\def\mIIAdof{$753$}			
\def\mIIAredc{$1.07$}			


\def\pmIAcnorm{$1.097^a$}			
\def\pmIAcdelgam{$0.01^a$}			
\def\pmIAnh{$6.5^{+0.1}_{-0.2}$}		
\def\pmIAga{$2.804^{+0.002}_{-0.001}$}			
\def\pmIApln{$49.69^{+0.07}_{-0.14}$}			
\def\pmIAq{$4.23^{+0.16}_{-0.08}$}			
\def\pmIAa{$0.93^{+0.01}_{-0.02}$}			
\def\pmIAi{$60^{+3}_{-1}$}			
\def\pmIAHden{$0.01^{+0.00060}_{-0.00002}$}			
\def\pmIAbbkT{$0.3$}			
\def\pmIAillum{$0.01^{+0.002}$}			
\def\pmIAren{$49.6^{+0.5}_{-0.4}$}			
\def\pmIAgn{$0.0047^{+0.0008}_{-0.0009}$}			
\def\pmIAge{$6.49 \pm 0.09$}			
\def\pmIAchi{$424$}			
\def\pmIAdof{$68$}			
\def\pmIAredc{$7.3$}			

\def\pmIIAcnorm{$1.097^a$}			
\def\pmIIAcdelgam{$0.01^a$}			
\def\pmIIAnh{$<0.2$}		
\def\pmIIAga{$2.31^{+0.05}_{-0.03}$}			
\def\pmIIAkTe{$16^{+5}_{-2}$}			
\def\pmIIAnthn{$5^{+1}_{-2}$}			
\def\pmIIAkTbb{$0.73^{+0.09}_{-0.04}$}			
\def\pmIIAdin{$6100^{+200}_{-300}$}			
\def\pmIIAa{$0.5^a$}			
\def\pmIIAq{$3^a$}			
\def\pmIIAi{$40^{+4}_{-2}$}			
\def\pmIIAreAfe{$1^a$}			
\def\pmIIArexi{$3.333^{+0.006}_{-0.027}$}			
\def\pmIIAren{$0.031^{+0.001}_{-0.002}$}			
\def\pmIIAgn{$0.005^{+0.001}_{-0.001}$}			
\def\pmIIAge{$6.20^{+0.13}_{-0.06}$}			
\def\pmIIAchi{$98$}			
\def\pmIIAdof{$68$}			
\def\pmIIAredc{$1.3$}			

\begin{table*}
\begin{center}
\small
\caption{Best fit parameters for fits of Models 1 and 2 separately to the \asca-GIS and \rxte-PCA September 23 1998 observations of \j1550. All limits are shown at 90\% confidence. Parameter descriptions: $N_{crabcorr}$ = Crab-correction to normalization, $\Delta\Gamma_{crabcorr}$ = Crab-correction to slope, $N_\mathrm{H}$ = Galactic hydrogen column density, $\Gamma$ = IC photon index,  $kT_\mathrm{e}$ = coronal electron temperature, $N_\mathrm{pow}$ = normalization of power-law component, $kT_\mathrm{bb}$ = disk blackbody temperature, $N_\mathrm{dbb}$ = disk normalization, $E_{\mathrm line}$ = energy of Gaussian emission line, $N_\mathrm{line}$ = normalization of Gaussian emission line, $q$ = emissivity profile index, $a_\mathrm{\star}$ = black hole spin, $i^{\circ}$ = disk inclination,  $H_\mathrm{den}$ = hydrogen number density in the top layer of disk, $F_\mathrm{Illum}/F_\mathrm{BB}$ = ratio of power-law to blackbody flux irradiating the disk, $A_\mathrm{Fe}$ = iron abundance, $\log{\mathrm{\xi}}$ = disk ionization, $N_\mathrm{refl}$ = normalization of reflection component, $\chi^2$ = total chi-squared, $\nu$ = degrees of freedom, $\chi^2_{\nu}$ = reduced chi-squared.}
\begin{tabular}{lcccc}
\hline
Parameters & Model~1 GIS & Model~1 PCA & Model~2 GIS & Model~2 PCA \\
& {\tt pow+gau} & {\tt pow + gau} & {\tt nthComp + diskbb} &  {\tt nthComp + diskbb} \\
& {\tt + kerrconv * refbhb} & {\tt + kerrconv * refbhb} & {\tt + gau + relxillCp} & {\tt + gau + relxillCp} \\
\hline
$N_{crabcorr}$ & \mIAcnorm\ & \pmIAcnorm\ & \mIIAcnorm\ & \pmIIAcnorm\ \\
$\Delta\Gamma_{crabcorr}$ & \mIAcdelgam\ & \pmIAcdelgam\ & \mIIAcdelgam\ & \pmIIAcdelgam\ \\
$N_\mathrm{H}$ ($10^{22}$~cm$^{-2}$) & \mIAnhI\ & \pmIAnh\ & \mIIAnhI\ & \pmIIAnh\  \\
$\Gamma$ & \mIAga\ & \pmIAga\ & \mIIAga\ & \pmIIAga\ \\
$kT_\mathrm{e}$ & \nodata\ & \nodata\ & \mIIAkTe\ & \pmIIAkTe\ \\
$N_\mathrm{pow}$ & \mIApln\ & \pmIApln\ & \mIIAnthn\ & \pmIIAnthn\ \\
$kT_\mathrm{bb}$ (keV) & \mIAbbkT\ & \pmIAbbkT\ & \pmIIAkTbb\ & \pmIIAkTbb\ \\
$N_\mathrm{dbb}$ & \nodata\ & \nodata\ & \mIIAdin\ & \pmIIAdin\ \\
$E_{\mathrm line}$ & \mIAge\ & \pmIAge\ & \mIIAge\ & \pmIIAge\  \\
$N_\mathrm{line}$ & \mIAgn & \pmIAgn & \mIIAgn & \pmIIAgn\  \\
$q$ & \mIAq\ & \pmIAq\ & \mIIAq\ & \pmIIAq\  \\
$a_\mathrm{\star}$ & \mIAa\ & \pmIAa\ & \mIIAa\ & \pmIIAa\ \\
$i^{\circ}$ & \mIAi\ & \pmIAi\ & \mIIAi\ & \pmIIAi\ \\
$H_\mathrm{den}$ ($10^{22}$~cm$^{-3}$) & \mIAHden\ & \pmIAHden\ & \nodata & \nodata  \\
$F_\mathrm{Illum}/F_\mathrm{BB}$ & \mIAillum\ & \pmIAillum\ & \nodata & \nodata \\
$A_\mathrm{Fe}$ & \nodata & \nodata & \mIIAreAfe & \pmIIAreAfe\ \\
$\log\xi$ & \nodata & \nodata & \mIIArexi\ & \pmIIArexi\ \\
$N_\mathrm{refl}$ & \mIAren & \pmIAren & \mIIAren\ & \pmIIAren\ \\
\hline
$\chi^2$ & \mIAchi\ & \pmIAchi\ & \mIIAchi\ & \pmIIAchi\  \\
$\nu$  & \mIAdof\ & \pmIAdof\ & \mIIAdof\ & \pmIIAdof \\
$\chi_{\nu}^2$  & \mIAredc\ & \pmIAredc\ & \mIIAredc\ & \pmIIAredc\ \\
\hline
\end{tabular}
\label{tab:indfits}
\end{center}
\end{table*}

\subsection{ \rxte\ }
During September 23--24 1998 a simultaneous, roughly 3~ks observation of \j1550 was taken with \rxte, and archival PCA and HEXTE (cluster A and B) data are publicly available. We extracted data from all PCUs of the PCA detector, and from both HEXTE clusters, discarding data within 10\,min from the South Atlantic Anomaly (SAA). For the PCA, we focus on PCU~2 alone, given its superior calibration over the other 4 PCUs. The PCU~2 spectrum has been corrected using the tool {\tt pcacorr} \citep{Garcia2014b} and $0.1$\% systematics have been added to all channels accordingly. We then ignore the PCU~2 counts in channels 1--4 and above $45$~keV. This limits the spectrum to roughly the 3--45~keV range. The HEXTE spectrum is also included, clusters A and B. The HEXTE B spectrum is corrected for instrumental effects using the tool {\tt hexBcorr} \citep{Garcia2016} analogously to the corrections made to the PCU~2 spectrum. We group both HEXTE spectra by factors of 2, 3, and 4 in the 20--30, 30--40, and 40--250 keV ranges respectively in order to achieve an oversampling of $\sim3$ times the instrumental resolution, and group at a signal-to-noise ratio of 4.  We constrain the HEXTE spectra to 25--200~keV after noticing the 20--25~keV region shows a spectral turnover at odds with the PCA data.

%
%
%
\section{Individual modeling}
\label{sec:imod}
\indent As discussed in Section~\ref{sec:intro}, the September 23--24 1998 \asca\ and \rxte\ spectra have been modeled with multiple representative models of an accretion disk, inverse Compton (IC) scattering corona, and reflection. We explore two separate reflection models and test the differences (qualitative and quantitative) when applied to these \asca-GIS and \rxte-PCA observations separately. We use {\tt XSPEC v.12.10.0c} for all our analysis. The models are as follows:

\begin{itemize}
\item {\bf Model~1:} \\
{\tt crabcorr*phabs*constant* \\ (powerlaw+kerrconv*refbhb+gaussian)};

\item {\bf Model~2:} \\
{\tt crabcorr*phabs*constant* \\ (nthComp+diskbb+relxillCp+gaussian)};
\end{itemize}

{\tt crabcorr} \citep{Steiner2010} is a corrective model which standardizes the detector response of a given instrument to retrieve the normalizations and power-law slopes of the Crab, based on the results of \cite{Toor1974}. For now we adopt the less-developed {\tt phabs} model for interstellar absorption since we are adopting the older interstellar elemental abundances of \cite{Anders1989} and cross-sections of \cite{BC1992} in accordance with the approach of \cite{Steiner2011}. We allow the Galactic hydrogen column density, $N_\mathrm{H}$, to vary freely, thus allowing us to test our results quantitatively against those of \cite{Steiner2011}. The model {\tt diskbb} \citep{Mitsuda1984} is a multi-temperature disk blackbody component, previously found to be dominant at soft energies in the hard-intermediate state of \j1550, constrained to a temperature of $\sim0.5$--$0.6$~keV (e.g., \citealt{Miller2003,Rodriguez2003,Gierlinski2003,Steiner2011,Hjalmarsdotter2016}). We introduced the {\tt kerrconv} and {\tt refbhb} models in Section~\ref{sec:intro}, these combine to model a reflection spectrum that includes a blackbody component to represent the disk emission as an irradiative component, as well as the relativistic blurring. {\tt RelxillCp} is a flavor of the {\tt relxill} suite of models \citep{Dauser2014,Garcia2014} which adopts a coronal Comptonization model as its irradiating continuum. The model {\tt nthComp} \citep{Zdziarski1996,Zycki1999} is that irradiating continuum, an IC model characterized by an electron temperature and power law index, in which the {\tt diskbb} component provides the input photon distribution for scattering. 

Model 1 is identical to that adopted by \cite{Steiner2011} in modeling of the same \asca-GIS data, albeit they applied this model to the GIS 2 and GIS 3 data separately, whereas here we have combined those spectra into an averaged GIS spectrum. Nonetheless, we find a consistent fit of this model to the GIS spectrum, with the model parameters and their confidence limits shown in the first column of Table~\ref{tab:indfits}. We allowed the inclination parameter of \texttt{refbhb} to extend to higher values than \cite{Steiner2011} and as a result find a slightly higher value, showing that indeed the application of this reflection model to the GIS spectrum naturally leads to high inferred disk inclination, and is consistent with being roughly aligned with the orbit ($i=84\pm3^{\circ}$). The unfolded spectrum and model, along with standardized $\chi$~residuals ($\chi=({\rm D}-{\rm M})/\Delta$, where ${\rm D}$ are the data counts, ${\rm M}$ are the model counts, and $\Delta$ are the data uncertainties), is shown in Figure~\ref{fig:indfits}. 

Model 2 is physically and geometrically similar to Model 1, the key difference being that the reflection component is described instead by {\tt relxillCp}, which does not include the disk blackbody emission as an irradiative component. Thus, we also must include the disk component explicitly with {\tt diskbb}. Both models make use of a simple Gaussian to fit the narrow Fe K line. The motivation for this comparison of Model 1 with Model 2 is to test the capability of each flavor of reflection model to model the \asca\ and \rxte\ data simultaneously, which thus far has not been performed on these strictly simultaneous observations. We focused particularly on whether there are distinct physical contrasts implied by the parameter constraints, and whether our results may differ from those of \cite{Steiner2011} in particular. 

The application of both Model 1 and 2 to the \asca-GIS and \rxte-PCA are shown in both Table~\ref{tab:indfits} and Figure~\ref{fig:indfits}. We find lower values for disk inclination when applying Model~1 to the PCA spectrum, and most notably, we struggle to fit this model successfully to that spectrum (best $\chi^2_{\nu}\sim7$). One notices instantly that when applying \texttt{refbhb} to a spectrum which extends to the higher energies covered by the PCA ($\sim45$~keV), the model meanders into an area of parameter space that plainly disagrees with the low-energy coverage of \asca\, and also struggles to fit to the Compton hump and apparent spectral turnover seen in the PCA spectrum. Instead we find that the component \texttt{relxillCp} included in Model~2 successfully fits to both the GIS and PCA spectra, and both fits lead to much lower constraints on the disk inclination ($i=23^{+8}_{-20}$ and $40^{+4}_{-2}$ for the GIS and PCA respectively). We note that the application of Model~2 to the GIS data alone, covering only 1--10~keV, implies total dominance of the reflection component over the irradiating coronal Comptonization spectrum (see Figure~\ref{fig:indfits}, top right panel): this is simply a result of the lack of broad spectral coverage which naturally limits the flux of the reflection component. One can see this discerning modeling behavior in the bottom right panel of Figure~\ref{fig:indfits}, whereby the 3--45~keV PCA spectrum, which includes the curvature modeled by the Compton reflection hump, as well as signs of a spectral turnover at high energies, leads to lower reflection fraction, with a dominant coronal IC component. 

\section{Simultaneous \asca\ and \rxte\ modeling}
\label{sec:smod}
We move on to fitting simultaneously to both the PCA (PCU 2) and \asca\ (GIS 2 and 3 combined) spectra. First we note that there appears to be a discrepancy between these two spectra, both in flux and photon index. The {\tt crabcorr} $\Delta\Gamma$ and $N$ estimates for each instrument (specifically GIS 2 on \asca\ and the PCA), based on extensive modeling of the Crab, are given by $\Delta\Gamma$ = -0.01, $N=0.97$ for \asca\ GIS 2 and $\Delta\Gamma=0.01$, $N=1.097$ for the PCA \citep{Steiner2010}, and we apply identical offsets to the combined GIS 2 and GIS 3 spectrum as those derived for GIS 2. Thus we should expect any successful model to fit to both instruments with these parameters fixed. We test this hypothesis by first modeling both datasets simultaneously (the combined GIS spectrum and the PCU~2 data), with a progression of simple-to-complex models, beginning with a simple absorbed power law, all the way to a model which includes  a corona, a disk, a relativistic reflection component, and a narrow distant reflection component. We move onto using the more recent absorption routine \texttt{TBabs} for completeness, with the latest interstellar medium abundances \citep{Wilms2000} and atomic cross section data \citep{Verner1996}, and fix the Hydrogen column density to $N_{\rm H}=10^{22}~\mathrm{cm^{-2}}$ in accordance with our model fits to the \asca\ GIS spectrum, as well as constraints from Galactic HI surveys \citep{Kalberla2005}; later in Section~\ref{sec:finalfit} we fit for $N_{\rm H}$, so we just fix the value for these initial tests. Figure~\ref{fig:gis_pca_offset} shows the ratio residuals of five separate model fits to the GIS and PCU~2 spectra, in which one can see that an offset in the power-law slope around 3--10~keV persists throughout the model progression. The results of the previous section show that Model 2 (equivalent to the final model fit shown in Figure~\ref{fig:gis_pca_offset}) fits well to both spectra individually. Therefore the persistence of an offset at energies $\sim$3--10~keV between the GIS and PCU~2 spectral fits is possible evidence for a slope/calibration difference. We suggest two possible explanations, and therefore two corresponding solutions to this mismatch: the GIS and PCA instruments suffer a cross-calibration error, or, this particular observation saw either of one of the instruments experience an energy gain shift. 

The mismatch between the GIS and PCU~2 spectra can be mitigated either by applying a gain shift correction to the GIS response (which could be expected given small gain shifts are seen in the GIS response for count rates exceeding $\sim100$~cts~${\rm s^{-1}}$, \citealt{Makishima1996}), or by applying a cross-calibration slope offset to the PCU~2 data. Figure~\ref{fig:gis_pca_corrected} shows two fits to the same data with Model~2, one in which we allow the $\Delta\Gamma$ parameter, as applied to the PCU~2 spectrum, to vary freely, and one in which we apply a gain shift to the GIS data. Both methods succeed in removing the majority of the low-energy residuals, thus circumventing the disconnect in spectral shapes between the two datasets. The slope and offset shifts in gain applied to the response of the GIS detector is slope = 1.02, offset = -0.04, thus on the order of a 2\% gain shift. The \texttt{crabcorr} slope offset applied to the PCA data to correct the slope offset between the GIS and PCA spectra is $\Delta\Gamma=0.10$, with a normalization of the flux with respect to the Crab of 1.37. One fundamental difference to the model parameters is the coronal temperature, which is a factor of a few higher when applying a gain correction. We also tested applying a gain correction to the PCA data, which we do not show here, and found this to be unsuccessful, yielding a slope and offset very close to the default. Though we not present the analysis here, we have also checked the effects of pileup in the PCA, and find that whilst there my be some change to the PCA slope due to pileup, it is minimal in comparison to the offset between the GIS and PCA. Given the relative success of applying a slope correction to the PCA in minimizing the residuals in the 3--10~keV range, we adopt this method in our final analysis of the data, as opposed to applying any corrections to the gains of either instrument response.

\def\mIIIcnormGIS{$0.97^a$}			
\def\mIIIcnormPCA{$1.23^{+0.01}_{-0.01}$}			
\def\mIIIcnormHXTA{$1.19^{+0.02}_{-0.02}$}			
\def\mIIIcnormHXTB{$1.19^{+0.02}_{-0.02}$}			
\def\mIIIcdelgamGIS{$-0.01^a$}			
\def\mIIIcdelgamRXTE{$0.090^{+0.006}_{-0.006}$}			
\def\mIIInh{$0.928^{+0.007}_{-0.009}$}		
\def\mIIIga{$2.13^{+0.01}_{-0.01}$}			
\def\mIIIkTe{$5.7^{+0.3}_{-0.2}$}			
\def\mIIInthn{$3.8^{+0.2}_{-0.2}$}			
\def\mIIIkTbb{$0.625^{+0.004}_{-0.007}$}			
\def\mIIIdin{$7100^{+400}_{-200}$}			
\def\mIIIa{$0.5^a$}			
\def\mIIIq{$3^a$}			
\def\mIIIi{$39.2^{+0.9}_{-0.9}$}			
\def\mIIIreAfe{$3.9^{+1.8}_{-0.6}$}			
\def\mIIIrexi{$4.44^{+0.21}_{-0.07}$}			
\def\mIIIren{$0.027^{+0.002}_{-0.001}$}			
\def\mIIIxin{$0.021^{+0.003}_{-0.004}$}			
\def\mIIIplgam{$1.7^a$}			
\def\mIIIplnorm{$0.22^{+0.04}_{-0.03}$}			
\def\mIIIabs{$20^a$}			
\def\mIIIchi{$1075$}			
\def\mIIIdof{$878$}			
\def\mIIIredc{$1.22$}			

\begin{table}
\begin{center}
\small
\caption{Median parameter values for fits of Model 3 to the ASCA and PCA/HEXTE September 23 1998 observations of \j1550, calculated from the final posterior probability distributions resulting from the MCMC chain. All limits are shown at 90\% confidence. Parameter descriptions: $N_{crabcorr}$ = Crab-correction to normalization, shown for each detector, $\Delta\Gamma_{crabcorr}$ = Crab-correction to slope, shown for each detector, $N_\mathrm{H}$ = Galactic hydrogen column density, $\Gamma$ = IC photon index,  $kT_\mathrm{e}$ = coronal electron temperature, $N_\mathrm{nth}$ = normalization of {\tt nthComp}, $kT_\mathrm{BB}$ = disk blackbody temperature, $N_\mathrm{dbb}$ = disk normalization, $q$ = emissivity profile index, $a_\mathrm{\star}$ = black hole spin, $i^{\circ}$ = disk inclination,  $A_\mathrm{Fe}$ = iron abundance, $\log{\mathrm{\xi}}$ = disk ionization, $N_\mathrm{rel}$ = normalization of {\tt relxillCp}, $N_\mathrm{xil}$ = normalization of {\tt xillverCp}, $\Gamma_\mathrm{pl}$ = photon index of high-energy tail, $N_\mathrm{pl}$ = normalization of high-energy tail, $E_\mathrm{expabs}$ = low-energy cutoff of high-energy tail, $\chi^2$ = total chi-squared, $\nu$ = degrees of freedom, $\chi^2_{\nu}$ = reduced chi-squared.}
\begin{tabular}{lr}
\hline
Parameters & Model~3 \\
&  {\tt crabcorr*TBabs(nthComp+diskbb} \\
& {\tt+relxillCp+xillverCp+(pl*expabs))}\\
\hline
$N_{crabcorr, GIS}$ & \mIIIcnormGIS\ \\
$N_{crabcorr, PCA}$ & \mIIIcnormPCA\ \\
$N_{crabcorr, HEXTE A}$ & \mIIIcnormHXTA\ \\
$N_{crabcorr, HEXTE B}$ & \mIIIcnormHXTB\ \\
$\Delta\Gamma_{crabcorr, GIS}$ & \mIIIcdelgamGIS\ \\
$\Delta\Gamma_{crabcorr, RXTE}$ & \mIIIcdelgamRXTE\ \\
$N_\mathrm{H}$ ($10^{22}$~cm$^{-2}$) & \mIIInh\ \\
$\Gamma$ & \mIIIga\  \\
$kT_\mathrm{e}$ & \mIIIkTe\ \\
$N_\mathrm{nth}$ & \mIIInthn\ \\
$kT_\mathrm{bb}$ (keV) & \mIIIkTbb\  \\
$N_\mathrm{dbb}$ & \mIIIdin\ \\
$q$ & \mIIIq\  \\
$a_\mathrm{\star}$ & \mIIIa\  \\
$i^{\circ}$ & \mIIIi\  \\
$A_\mathrm{Fe}$ & \mIIIreAfe\  \\
$\log{\mathrm{\xi}}$&  \mIIIrexi\ \\
$N_\mathrm{rel}$ & \mIIIren  \\
$N_\mathrm{xil}$ & \mIIIxin  \\
$\Gamma_\mathrm{pl}$ & \mIIIplgam\ \\
$N_\mathrm{pl}$ & \mIIIplnorm\ \\
$E_\mathrm{expabs}$ & \mIIIabs\ \\
\hline
$\chi^2$ & \mIIIchi\   \\
$\nu$  & \mIIIdof\  \\
$\chi_{\nu}^2$  & \mIIIredc\  \\
\hline
\end{tabular}
\tablenotetext{a}{Fixed parameter}
\label{tab:finalfit}
\end{center}
\end{table}

\begin{figure*}[ht!]
\includegraphics[width=\linewidth]{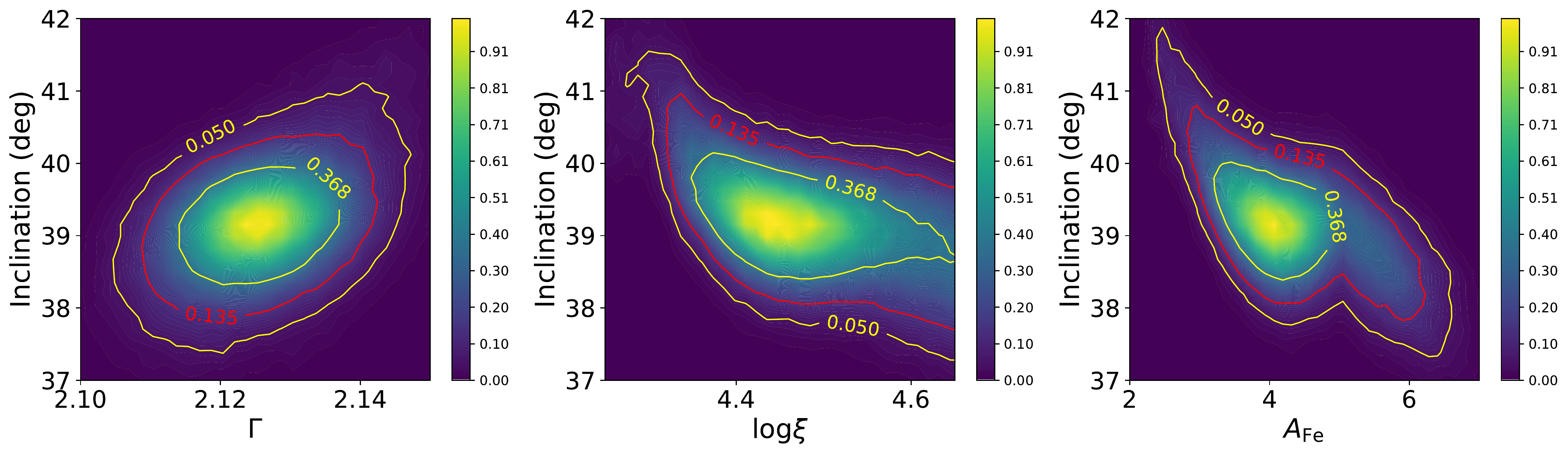}
\caption{2D contours of the disk inclination and three key model parameters associated with the iron line region of the X-ray reflection spectrum: the photon index of the coronal-IC scattering region, $\Gamma$, the disk ionization, $\log\xi$, and the iron abundance, $A_\mathrm{Fe}$.The side bar shows the probability density scale, and contours at $1\sigma$, $2\sigma$, and $3\sigma$ are shown with yellow, red, and gold lines respectively.  }
\label{fig:contours}
\end{figure*}

\begin{figure}[ht!]
\includegraphics[width=\linewidth]{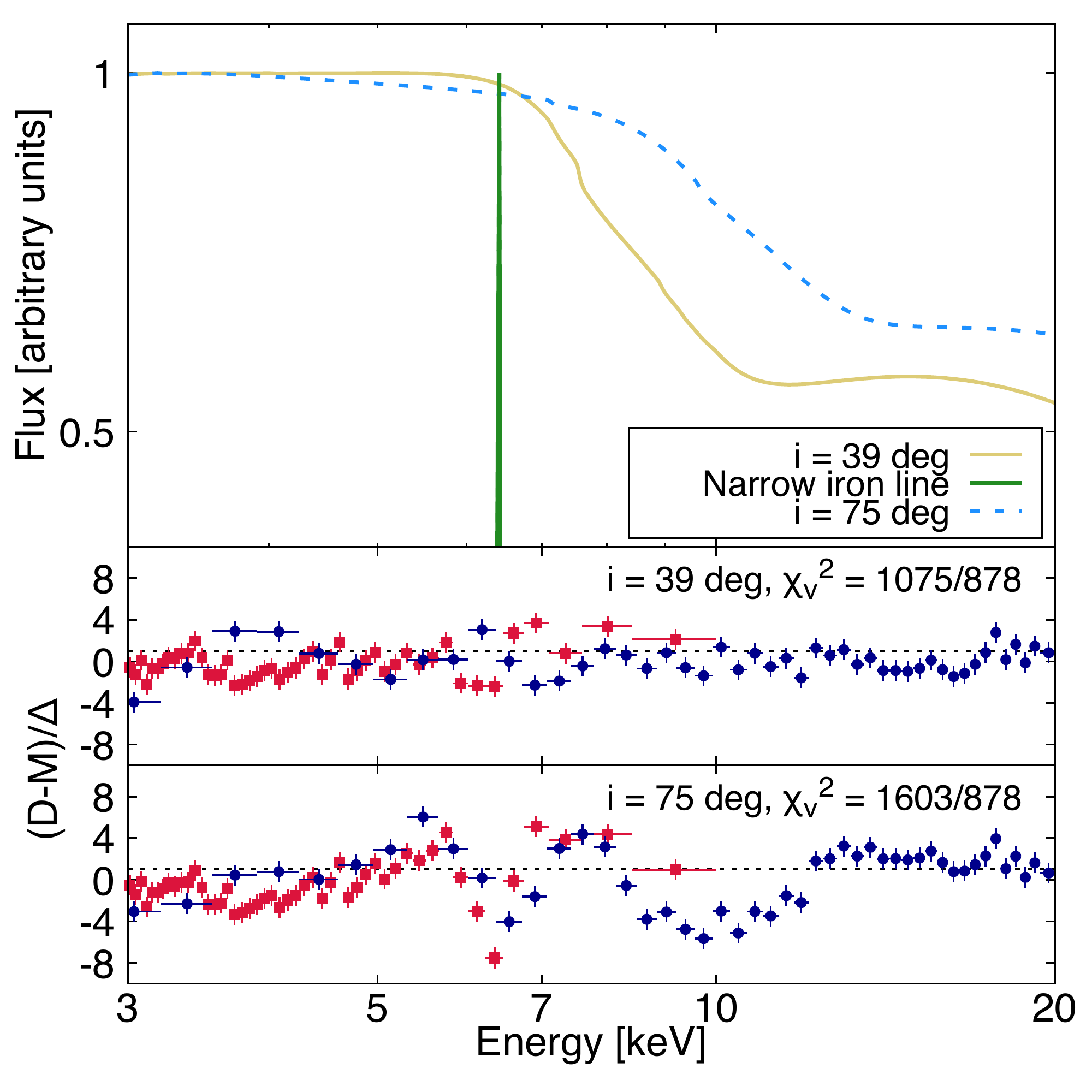}
\caption{A comparison of the spectrum given by Model~3 in which the inclination parameter is adjusted from the best fit value of $\sim39^{\circ}$ to the orbital inclination at $\sim75^{\circ}$. The top panel shows the model reflection spectra in arbitrary flux units, with the narrow reflection component of our best fit model shown for perspective, and the bottom panels show the $\chi$ residuals of the fit of Model~3 with $i=39^{\circ}$, the best fit value, and $i=75^{\circ}$, a fixed value. Red squares show the \asca-GIS residuals, blue points show the \rxte-PCA residuals. }
\label{fig:incl}
\end{figure}


\subsection{Final fit}
Adopting a slope correction between the GIS and PCA instruments, we now fold in the \rxte-HEXTE data and apply our final model:
\label{sec:finalfit}

\begin{itemize}
\item Model~3:\\
\texttt{crabcorr*TBabs(nthComp+diskbb+relxillCp}\\
\texttt{+xillverCp+(powerlaw*expabs))}
\end{itemize}

We fit Model~3 to the full broadband 1--200~keV spectrum. Here we have replaced the Gaussian model, previously used to take account of the additional narrow-line residuals in the iron-line region, with the \texttt{xillverCp} model. This represents a distant reflection component, undistorted by the general-relativistic effects associated with a reflected component lying close to the black hole horizon (accounted for by \texttt{relxillCp}). We tie the iron abundance of \texttt{relxillCp} and \texttt{xillverCp} together since both reflectors are components in the same accretion disk, and we fix the log ionization of the distant reflector (\texttt{xillverCp}) to $\log\xi=0$ (i.e., an almost neutral gas), since the irradiating flux impinging on the distant reflector is expected to be orders of magnitude lower than that which strikes the inner regions of the disk. The ionization of the relativistic reflection component (\texttt{relxillCp}) is a free parameter. We continue to fix the black hole spin to $a_\mathrm{\star}=0.5$ \citep{Steiner2011}, given the lack of strong constraints on its value when allowed to vary freely (for example, see column 2 of Table~\ref{tab:indfits}). As such the disk inner radius is also kept fixed at the innermost stable circular orbit (ISCO). The temperature of the photon distribution impinging on the corona is set to the disk temperature of the {\tt diskbb} component ($kT_{\rm BB}$), and $\Gamma$, the photon index, is tied between the {\tt nthComp}, {\tt relxillCp}, and {\tt xillverCp} components. 

In addition to the IC component and two reflection components, we also include an additional high-energy tail to the model, with a low-energy cutoff at 20~keV. This component represents the evidence for non-thermal electrons in a hybrid plasma, leading to a non-thermal IC tail, making our model similar to that adopted by \cite{Hjalmarsdotter2016}. This high-energy component is required by the HEXTE spectra with a low statistical significance, and we include it mostly as a nuisance component, remaining indifferent to the physical implications. As such we fix the index, $\Gamma_\mathrm{pl}=1.7$, in accordance both with standard diffusive shock acceleration theory (see, e.g., \citealt{Drury1983}), and with the slope of the injected non-thermal particle distribution in the modeling performed by \cite{Hjalmarsdotter2016}. We cut the low-energy portion of the high-energy tail off at 20~keV to ensure limited degeneracy with the components modeling the iron line region, since this additional power law is unconstrained in the lower energy regions. The choice of 20~keV is an appropriate one because it allows us to model out the high-energy tail, but it does not introduce confusion in the Fe K line region. Since the addition of this component clearly means the cut-off energy of the thermal pool of electrons in the corona does not properly describe the limiting photon energy of the disk irradiator for reflection, we also fix the electron temperature in \texttt{relxillCp} and \texttt{xillverCp} to 200~keV. As discussed by \cite{Garcia2015_2}, the cutoff energy for reflection can be constrained by, and thus has a strong impact on, the lower-energy portion of the reflection spectrum. It is therefore important to consider the high-energy radiation and make sure the reflection model knows about it. Our choice to fix the cutoff at higher energies than given by the temperature of thermal IC component does have a drawback, since the high-energy tail has a harder spectral slope than the {\tt nthComp} component, however, this choice is the most consistent of the two. 

We first fit the full spectrum and run an error analysis, all using \texttt{XSPEC v.12.10.0c} \citep{Arnaud1996}, and then make use of the \texttt{EMCEE} \citep{Foreman-Mackey2013} Markov-Chain Monte Carlo (MCMC) routine with a simple {\tt PYXSPEC} wrapper to better explore the parameter space of the model. In this way we can search for any possible multi-modality to the fit, with particular attention paid to possible correlations between the disk inclination and other model parameters. We initialize 100 walkers per free parameter and run the MCMC exploration for one million steps (for every walker), and subsequently burn the first 30\% of the run to ensure the final distributions are fully converged with limited noise. We place flat log prior distributions on all normalization parameters, in order to ensure sensible walker step sizes in the MCMC chains. 

Figure~\ref{fig:finalfit} shows the final model fit to the full 1--200~keV \asca\ and \rxte\ spectrum, based on our best fit using the standard Levenberg-Marquardt algorithm in \texttt{XSPEC}, and Table~\ref{tab:finalfit} shows the final parameters and their 90\% confidence limits after running the full MCMC routine. Most strikingly we find, in accordance with the results of fitting only the PCA data with Model~2, that the disk inclination is constrained to a much lower value ($i=39.2^{\circ\, +0.9}_{\,\,\, -0.9}$) than given by the orbital inclination measurement of $75^{\circ}\pm4$. Figure~\ref{fig:contours} shows 2D contours which demonstrate correlations (or lack there of) between the disk inclination $i$ and other key parameters. There is evidence for a weak positive correlation between $i$ and the coronal photon index, $\Gamma$, as well as anti-correlations between $i$ and the disk ionization ($\log\xi$) and iron abundance ($A_\mathrm{Fe}$), but all these key parameters are well constrained, such that we can safely conclude the inclination of the disk must be low. Figure~\ref{fig:corner} in the Appendix (Section~\ref{sec:app}) shows the full 1D posterior distributions of every free parameter of the model, as well as all the 2D contours which show potential parameter correlations. One can see that there are not many clear correlations between our parameters. 

The reason for such a strict constraint on the inclination is geometrical in nature, and a result of the relativistic Doppler broadening of the spectral reflection features, in particular the Fe K emission. Figure~\ref{fig:incl} demonstrates the effect of raising the inclination parameter in the model component \texttt{relxillCp} to $75^{\circ}$. The more the disk is inclined, the more the blue wing of the iron line increases in flux due to Doppler broadening, dramatically altering the shape of the spectrum in the $\sim6$--$10$~keV region. Given the high count rates of the source in this bright hard-intermediate state, the data plainly rule out a high-inclination relativistic reflection component. 

For completeness we also perform the following tests. First we fit Model 3 with both the inner disk radius ($R_{\rm in}$) and emissivity index ($q$) free to vary. We do this in order to test for additional degeneracies with inclination that may be associated with the relativistic effects both parameters generate in the Fe K line region. We present a brief description of these results in the appendix (Section~\ref{sec:app2}), but to summarize, neither $R_{\rm in}$ nor $q$ can be constrained, both are consistent with the fixed values previously chosen ($R_{\rm in}=R_{\rm ISCO}$, $q=3$). We also find that the inclination is still entirely consistent with the value shown in Table~\ref{tab:finalfit}. Second, we tested the effect of allowing both to vary freely with the inclination fixed at $75^{\circ}$, the orbital inclination. Again we find that $R_{\rm in}$ and $q$ are unconstrained at $3\sigma$ confidence, with $R_{\rm in}$ constrained to below a few times $R_{\rm ISCO}$ at $2\sigma$ confidence. We also obtain a best fit $\chi^2=1113/877=1.27$. We conclude from these tests that there is a strong preference, regardless of the freedom allowed in the fit of Model~3, for the inclination of the disk of \j1550\ to be low ($i\sim40^{\circ}$).  

We also find that the coronal electron temperature is very low ($kT_\mathrm{e}=5.8^{+0.3}_{-0.4}$), which agrees with the results of \cite{Hjalmarsdotter2016}, who applied various flavors of hybrid plasma models to spectral data taken within the same period with both \asca\ and \rxte. We note, however, that comparison of our modeling results with those of \cite{Gierlinski2003} and \cite{Hjalmarsdotter2016}---despite the concurrence in the observational state of \j1550\ and the instrumental data modeled---has its limitations due to the methods used to group data together. Both these works made use of co-added PCA spectra comprised of individual exposures taken within a roughly two-week period between September 23 and October 6, 1998. The authors justify this grouping of the data by arguing that the X-ray spectral hardness of \j1550\ does not vary significantly in this time frame. However, after careful inspection we found that at least for the purposes of reflection studies, the spectral hardening ($\Delta\Gamma\sim0.1$) in this window must be accounted for, particularly for data with Crab level count rates in the PCA detectors. In fact, the choice to group the spectra in this way actually hides the distinct disagreement in spectral slope between the \asca-GIS and \rxte-PCA observations, which coincidentally happens to be around an offset of 0.1 in the photon index. Our modeling manages to account for the cross-calibration differences between the \asca\ and \rxte\ data in a way which minimizes loss of interpretation of the physics.

\section{Discussion}
\label{sec:discussion}
The focal point of our modeling results is the alarmingly low disk inclination, $i\sim40^{\circ}$. There is a wealth of evidence accrued to date which show that not only is the orbital inclination of \j1550\ high ($\sim75^{\circ}$; \citealt{Orosz2011}), but the jet and counter-jet inclinations are correspondingly high ($\theta_{\rm jet}\sim71^{\circ}$; \citealt{Steiner2012}). Whilst, as we shall discuss in this section, the apparent alignment of the jet with the orbit, does not confirm that the spin is aligned with the orbit---the transient jet emission modeled by \cite{Steiner2012} originates from distances orders of magnitude beyond the inner $\sim10~r_{\rm g}$ within which the spin has an impact on the flow---it is nonetheless interesting that evidence for lower disk inclination is only observed in these reflection modeling results. 

In addition to direct attempts to characterize the orbit and jet axes, there are other correlations, typically seen in the X-ray variability of BHBs, which distinguish low inclination sources from high inclination sources. QPOs detected in the X-ray power spectra of BHB light curves have shown inclination-dependence, with evidence for type-B QPOs appearing stronger in low-inclination sources, and type-C stronger in high-inclination sources \citep{Motta2015}. \j1550\ fits with the latter of these two distributions. \cite{Heil2015b} also independently characterized this bimodal distribution through their novel classification of X-ray variability (power-colour hue). In addition, analysis of the phase lags of QPOs of multiple BHBs implies two distinct trends with inclination, with \j1550\ shown to fit in the high-inclination distribution \citep{vandenEijnden2017}. Indeed, QPOs are detected in the \rxte-PCA observation we have modeled in this work, and at multiple stages of the 1998/99 outburst of \j1550\ \citep{Remillard2002a}. 

QPOs are not the only BHB phenomenon which shows correlations with inclination. Lags of hard X-ray emission with respect to soft emission, a long known component of the X-ray variability of BHBs (e.g., \citealt{Miyamoto1988, Kazanas1997, Nowak1999}), also appear to show an inclination-dependence. \cite{Reig2019} recently found that a correlation between X-ray photon index ($\Gamma$) and these time lags shows increased scatter in higher-inclination BHBs. \j1550\ has large scatter in its time lag estimates as a function of $\Gamma$, in line with its high orbital inclination. 

Finally, the radio jets of BHBs also have trends with orbital inclination. \cite{Motta2018} found that high-inclination BHBs are more radio-quiet than low-inclination BHBs, again with \j1550\ being a part of the high-inclination group. 

All these clear indicators of the orientation of the binary and accretion flow of \j1550\ pose the question, why are we seeing a low inclination in the reflector? There are a few possible explanations for the mismatch, all of which we explore in the following sub-sections. Firstly, it is plausible, though seemingly unlikely, that the inner regions of the disk and jet are misaligned with the orbit, due to spin-orbit misalignment and the torquing of the inner flow, known as the Bardeen-Petterson effect \citep{Bardeen-Petterson1975}. Secondly, the geometries of the inner disk/corona/jet may all be quite different than is implied by our simplistic modeling of \j1550, i.e., the inner disk may not be a simple razor-thin Shakura-Sunyaev slab, as is assumed in the model {\tt relxillCp}. The jet/corona geometries may be more complex in this hard-intermediate state, such that they are outflowing components with large opening angle---this would alter the observed reflection spectrum. We now discuss this in more detail. 

\subsection{Warped disk?}
 The concept of misalignment between the inner and outer regions of an accretion flow goes back to the idea that viscous torques in the disk, along with Lense-Thirring precession \citep{Lense-Thirring} due to misalignment between the black hole spin and the disk plane, will cause the inner disk to line up with the spin axis \citep{Bardeen-Petterson1975}. The term warp arises due to the transition between an outer disk region which is aligned with the orbit, and an inner disk region aligned with the black hole spin axis. Thus, the extent of this warp depends on the extent of this misalignment between the spin axis and the orbital axis. Whilst therefore it is possible to detect misalignment between the inner disk (the inner $\sim10~r_{\rm g}$ region principally responsible for the relativistic reflection component in the spectrum) and the outer disk, or orbit, the likelihood of this being true of \j1550, where we see a discrepancy of $>30^{\circ}$, depends on two key factors: (i) the probability of the binary system forming with such a large misalignment between the black hole spin axis and the binary plane, and (ii) the timescale for this misalignment to diminish due to accretion.
 
 Factor (i), the probability of a binary system forming with high spin-orbit misalignment, can be roughly predicted through a population synthesis study, such as the one conducted by \cite{Fragos2010}. In their work the authors showed that since those binaries which survive the BH-forming supernova explosion of one of the stellar components tend to be those with smaller `kicks', the resulting distribution of spin-orbit misalignments is highly skewed toward low angles, $<10^{\circ}$. However, this still leaves a small proportion of all Galactic BHBs that will have very high spin-orbit misalignments, even up to $90^{\circ}$. Thus, although we should expect a small misalignment between the spin and orbital axes in most BHBs, we cannot rule out the seemingly atypical $\sim30^{\circ}$--$40^{\circ}$ in \j1550\ implied by our reflection modeling results. 
 
 Factor (ii), the prediction that the process of accretion onto the black hole should result in the alignment of the spin and orbital axes in BHBs, has been estimated in multiple theoretical studies (e.g., \citealt{Natarajan1998,Maccarone2002, Martin2008}). \cite{Natarajan1998} calculated the timescale for the spin and accretion disk axes in active galactic nuclei (AGN) to align, based on earlier work by \cite{Papaloizou1983} showing that warps travel in the disk in a way that is governed by the internal hydrodynamics of the disk. This timescale, $t_\mathrm{align}$, is proportional to the black hole spin and the disk viscosity, and inversely proportional to both the source luminosity and black hole mass, though the mass dependence is very weak.
 
\cite{Maccarone2002} showed, based on this formalism, that black holes formed with relatively low spin have alignment timescales $\sim10\%$ of the binary lifetime, $t_\mathrm{bin}$. We can tune this formula to the specific case of \j1550, giving:
 
 \begin{equation}
 \label{eq:talign}
 \frac{t_\mathrm{align}}{t_\mathrm{bin}}=0.003~a_\mathrm{\star,0.5}^{11/16} \alpha_{0.03}^{13/8} L_\mathrm{0.1L_{E}}^{1/8} 
M_9^{15/16} \epsilon_{0.08}^{7/8} M_\mathrm{2, 0.3}^{-1} \,\, ,
 \end{equation}
 
 where $a_\mathrm{\star,0.5}$ is the black hole spin, assumed to be 0.5, $\alpha_{0.03}$ is the disk viscosity parameter, $L$ is the luminosity in units of 10\% $L_\mathrm{E}$, the Eddington luminosity, $M_9$ is the black hole mass tuned to $9~M_{\odot}$, $\epsilon_{0.08}$ is the accretion efficiency, assumed to be roughly 8\% in the case of \j1550, and $M_{2,0.3}$ is the mass of the donor star, determined to be $\sim0.3~M_{\odot}$ \citep{Orosz2011}. Adopting the more up-to-date analytical calculation of \cite{Martin2008} we find this fraction increases by roughly a factor of 2--3. Nonetheless, the spin axis should in principle align with the orbit within the binary lifetime, and one should not expect to see such a large misalignment. Indeed \cite{Steiner2012} used similar principles to support their finding that the spin axis (and thus jet launching axis) is aligned with the binary orbit in \j1550. 
 
 Regardless of the apparent likelihood that the inner disk and jet of \j1550\ are aligned with the binary orbit, the comparatively low measurement we have obtained from our reflection modeling naturally leads us to consider a geometrical scenario that could explain the disparity. The current breeding ground for physical estimates of the behavior of misaligned disks lies in computationally expensive general relativistic magnetohydrodyamic (GRMHD) simulations (e.g., \citealt{Fragile2007,Liska2018,Liska2019}) that can capture the fluid dynamics of a plasma with magnetic fields and strong gravity. One such code adopted to simulate black hole accretion, H-AMR, is now being applied to the spin-orbit misalignment problem, with the goal of characterizing the entire jet-corona-disk geometry \citep{Liska2018,Liska2019}. The results thus far indicate that in the very inner regions ($<10~r_\mathrm{g}$) the jet, disk and corona all align with the black hole spin axis. However, what is rather more noteworthy is that \cite{Liska2018} find the disk remains at least partially aligned with spin axis out to further distances than both the jet and corona, not approaching full alignment with the plane of the feeding torus until 10s of $r_\mathrm{g}$. An inner disk that is misaligned in this way would naturally lead to a lower inclination in the reflection modeling, since the strongest portions of the reflection spectrum occur in those inner regions.
 
 \cite{Liska2018} first investigated the effects of applying a relatively minor ($10^{\circ}$) spin-orbit misalignment, which is difficult to relate to the apparent $>30^{\circ}$ misalignment we find here for \j1550, whereas \cite{Liska2019} extended this work to the exploration of such large misalignments, as high as $\sim65^{\circ}$, and a much thinner disk ($H/R<0.03$). At such high misalignments and low disk thickness, the inner disk is seen to tear away from the broader accretion flow. The inner disk and jet both precess rapidly, with both the outer disk and jet aligning with the binary orbit. 
 
 Strong caveats arise when drawing any connection between our reflection modeling results and the more complex/involved simulations results of \cite{Liska2018} and \cite{Liska2019}. Such simulations tell us little about how radiative losses and a full ray-tracing treatment may impact the resultant geometry and thus also what the reflection spectrum may look like. In addition, GRMHD simulations are giving us insight into the physics of accretion in the presence of strong, dynamically important, magnetic fields. Furthermore, the misalignment of the jet/corona/disk system depends strongly on the assumed disk thickness, which is another element of the geometrical setup we do not consider in any detail in our analysis (the disk is approximated as a slab in the reflection model {\tt relxill}). Thus the details of how dynamically important magnetic fields may impact the geometry and thus what the reflector sees, as well as the observer, and the role of disk thickness on both the degree of misalignment and the resultant shape of the Fe K line in the reflection spectrum, are still entirely open questions. 
 
 Observational constraints on the jet inclination over time also seem to rule out a warped inner jet, at least. The extensive years-long radio observations of the transient jet formed by \j1550\ show no signs of precession \citep{Corbel2002b}. The 7-Crab flare which preceded the detection of the transient jet, and thus presumed to coincide with the launching window of the jet, lasted roughly 1 day, significantly longer than the timescale of the predicted precession of the inner $\sim10~r_\mathrm{g}$ of the accretion flow if it were misaligned with the orbit (basic size scale estimates give precession timescales orders of magnitude shorter than the day-long launching window, see, e.g., \citealt{Fragile2007,Ingram2009}). The precession of these inner regions should be observable as a significant angular variation in the outer jet, yet this is not observed. A good test case to compare to is the recent result concerning BHB V404~Cygni, in which high-resolution radio observations of its transient jet during a bright 2015 outburst reveals precession on minute timescales \citep{Miller-Jones2019}. 
 
 \subsection{Beyond the razor-thin disk}
 If the inner regions of the accretion flow, as implied by the wealth of evidence cited thus far, are aligned with the binary orbit, then we must invoke another scenario to explain the low inclination of the reflector. Another possible explanation is simply that the inner regions of the disk are not geometrically thin, as prescribed in detail by \cite{SS1973}. The luminosity of \j1550\ during September 23--24 1998, the window of these observations, reached $\sim0.1~L_\mathrm{Edd}$. At these luminosities we should not expect the disk to increase its scale height due to either radiative or thermal pressures (see, e.g., \citealt{SS1973, Paczynsky1980}), since such effects are only expected to occur at close to Eddington luminosities. However, it could be that the radial profile of the disk has some structure within the inner regions which is explained neither by high radiative flux, nor thermal pressures. Recent work by \cite{Jiang2019} has shown that simulations are now being used to explore the vertical structure of disks in more magnetically-dominated states, for example. The inclination determined by the reflection modeling may be somewhat of an artifact of a complex inner accretion flow, rather than a true representation of the orientation of the disk itself.

 An additional possibility may be that at high disk inclinations (assuming there is no warp in the disk) there is some obscuration of the blue-shifted line emission on the front-facing side of the disk. As shown in Figure~\ref{fig:incl}, at higher inclinations, the iron emission line is Doppler-broadened, and this emission originates in gas traveling towards the observer in front of the black hole. If this portion of the disk were to be obscured due to the scale-height profile, we should expect a reduction in the blue-ward flux of the line, much akin to an iron line predicted by reflection at lower disk inclinations. \cite{Taylor2018} have explored this scenario, and find that when the coronal height in a lamppost model is comparable to the inner disk thickness, one indeed expects modifications to the line profile due to obscuration effects, an inclination dependent effect. At this stage we make this suggestion speculatively given the limited work on the effects of disk geometries on reflection, and leave proper investigation to future work, since such a test requires substantial modifications to the \texttt{relxill} model to include these geometrical dependencies, and indeed no other model currently has such capabilities. 
 
 \subsection{Outflowing corona/jet}
 The location of the irradiating source for disk reflection determines the fraction of photons which strike the disk, as well as the location of the disk at which they strike, due to the relativistic effects in the presence of the strong gravity of the black hole. In addition, as shown by \cite{Dauser2013}, if the irradiating source is in motion at speeds comparable to the speed of light, similar implications arise. As shown in multiple papers over the past few decades, there is an inherent degeneracy between a static corona which produces the observable hard X-ray spectrum in BHBs, and a jet/outflow in relativistic motion \citep{Beloborodov1999,mnw05,Connors2017,Connors2019}. Furthermore, the recent simulations of \cite{Liska2018,Liska2019} have shown that the jet/corona/disk trinity is a connected physical system in which the individual components dynamically impact one another. For example, the jet may actually be collimated by the pressure provided by the enveloping corona. 
 
This leads to a natural question: How would the broad Fe K line profile appear if the irradiating source is a region travelling at mildly relativistic velocities perpendicular to the disk? \cite{Dauser2013} addressed this to some degree, and showed that indeed the height and velocity of the irradiating source (considered in this case to be a lamppost) strongly effects the broad line profile. However, these effects primarily occur within the red portion of the line, since the primary changes are in the gravitational redshift of the photons. Nonetheless, it is possible, if the source height is beyond the inner $\sim10~r_{\rm g}$ region, that the peak blueward emission due to Doppler effects may be reduced, since a greater source height leads to more photons striking the outer regions of the disk. Thus this is an additional effect which could lead one to derive a low disk inclination when performing reflection modeling of the kind we have performed in this paper, though it is difficult to generate such a large ($>30^{\circ}$) discrepancy with respect to the binary inclination with this one effect. 

As well as the location and velocity of the irradiating source, the lateral structure of the jet/corona may have an impact on the inclination estimate. \cite{Liska2019} investigated Bardeen-Petterson misalignment in accreting black holes with very thin accretion disks ($H/R<0.03$), and showed that in such cases the jet/corona are very broad geometrically. One would need to extend current reflection models to include such a scenario in which the irradiating source is not only vertically extended, but has lateral structure, such that the origin of much of the irradiating photons is not at $x=0$. 
  
 \subsection{High density disk?}
 Another significant result of our reflection modeling of \j1550\ is the high iron abundance ($A_\mathrm{Fe}=3.9^{+1.7}_{-0.5}$). High iron abundances have been measured for multiple BHBs and AGN alike \citep{Garcia2018b} in reflection modeling \citep{Parker2015,Fuerst2015,Garcia2015}, and the current interpretation of these estimates is that the disk density in reflection modeling has until recently been underestimated \citep{Tomsick2018,JJiang2019}. Thus given the very high iron abundance resulting from our modeling, we suggest future modeling of the reflection spectrum of \j1550\ ought to be tuned to higher disk densities. An increase in disk electron number density to beyond $\sim10^{20}~\mathrm{cm^{-3}}$ results in excess free-free heating via bremsstrahlung, which leads to higher soft X-ray flux. This increased soft X-ray flux can subsume the blackbody disk emission to a certain extent, and has even been invoked to explain the soft excess in AGN \citep{Garcia2019}. 
 
 As yet it is not entirely clear what the effects of introducing higher disk densities may be on the inclination constraint we have presented in this work. Given that the broadening of the blue wing of the iron line is largely inclination-dependent due to relativistic effects that are linked directly to the geometry of the system, it seems unlikely that higher disk densities will somehow alter the resultant geometrical interpretation of the system. Indeed recent applications of high-density reflection models to Cygnus~X-1 and GX~339$-$4 spectra have shown that the iron abundance is the only parameter which changes with respect to low-density modeling \citep{Tomsick2018,JJiang2019}, most notably the inclination remains unaffected. We have chosen not to explore high-density models in our fits because the models are currently limited in terms of the range of values of key physical parameters, such as for example the electron temperature (fixed at a single value in the most recent high-density models; \citealt{Garcia2016b}). 
 
 \subsection{Implications for continuum-fitting and spin}
 \cite{Steiner2011} provided two independent constraints on black hole spin via both modeling of the soft-state thermal continuum of \j1550, and through reflection modeling. Both methods yielded a low spin for the source, thus giving an estimate of $a_\mathrm{\star}\sim0.5$ as a an average of the two measurements. However, this estimate was guided by the assumption that the disk is inclined to the same degree as the binary orbit ($i\sim75^{\circ}$). \cite{Steiner2011} showed that the black hole spin is quite strongly anti-correlated with the disk inclination in thermal-continuum modeling. This leads to the obvious implication of our much lower inclination estimate, that adopting our inclination would produce a much higher value of spin from continuum fitting.
 
 \subsection{Other key parameter constraints}
\begin{figure*}
\includegraphics[width=\linewidth]{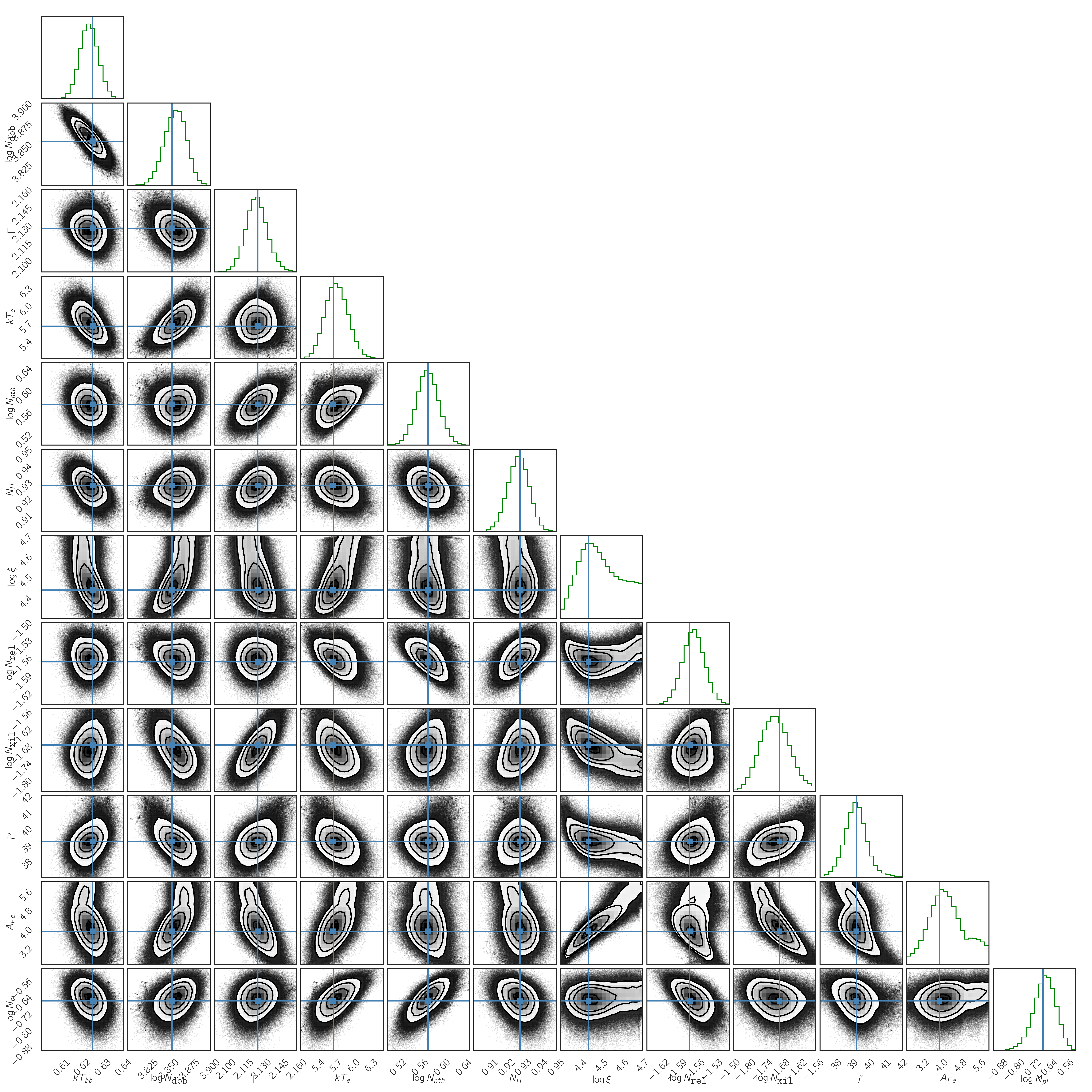}
\caption{Corner plot showing 1D and 2D distributions and contours resulting from MCMC parameter exploration of the fit of Model~3 to the simultaneous \asca-GIS, \rxte-PCA-and-HEXTE spectra taken on September 23--24 1998. The MCMC routine was run for $10^6$ steps with 100 walkers per free parameter, and the contours/distributions are generated after discarding the first 30\% of the chain. Blue lines/squares show the best fit parameters evaluated prior to running the MCMC chains, and are represented in the spectral fit shown in Figure~\ref{fig:finalfit}. Parameters shown are as follows: Inner disk temperature ($kT_{\rm bb}$), disk normalization ($\log N_{\rm dbb}$), IC photon index ($\Gamma$), coronal electron temperature ($kT_{\rm e}$), coronal normalization ($\log N_{\rm nth}$), Galactic hydrogen column density ($N_{\rm H}$), ionization of the reflector ($\log\xi$), {\tt relxillCp} normalization ($\log N_{\rm rel}$), {\tt xillverCp} normalization ($\log N_{\rm xil}$), disk inclination ($i$), iron abundance ($A_{\rm Fe}$), and normalization of the high-energy tail ($N_{\rm pl}$). }
\label{fig:corner}
\end{figure*}

 Much of our discussion has been focused thus far on the inclination constraints of our reflection modeling. There are however a few other interesting parameter constraints to discuss. Table~\ref{tab:finalfit} and Figure~\ref{fig:corner} show the best fit parameters with 90\% limits and the associated MCMC parameter correlations. One can see that the coronal electron temperature is very low ($kT_\mathrm{e}\sim6$~keV), a feature of hard-intermediate state BHBs (see, e.g., \citealt{rm06}), and in line with results of modeling \j1550\ with a hybrid coronal plasma \citep{Hjalmarsdotter2016}. We also found very high disk ionization ($>10^4~\mathrm{erg~cm~s^{-1}}$), and though its value is tightly correlated with the iron abundance in the disk, it is well constrained to such high values. In addition, though the ionization may appear surprisingly high, it can be expected for softer states in which the Fe K emission is stronger than in hard states in which we typically see lower values \citep{Garcia2013}. The constraints on the coronal power-law photon index ($\Gamma$), hydrogen column density ($N_\mathrm{H}$), and disk blackbody temperature ($kT_\mathrm{BB}$) all agree well with previous estimates \citep{Miller2003,Tomsick2003} of the spectral parameters of \j1550\ in its hard-intermediate state. In addition, we checked the $N_\mathrm{H}$ value is consistent with simple modeling of {\it Chandra} grating spectra during an observation in May 2000 when the source was in a similar hard-intermediate state, ObsID 680 and 681, finding only a difference of $\sim0.2$ between the derived value and our results from fitting Model 3 (a minor difference given the slight difference in model approach). There are, as already discussed in Section~\ref{sec:smod}, some differences in our spectral fitting results when compared to the modeling by \cite{Gierlinski2003,Hjalmarsdotter2016}. These differences are due to the differences in treatment of the data, with both the respective previous works combining PCA spectra across a small range in X-ray hardness. Selecting the data in this way, due to the higher PCA count rates being skewed toward harder spectra, actually serendipitously mitigates the offset in spectral slope between the \asca-GIS and \rxte-PCA data, a feature that we have shown is important to account for in the September 23--24 1998 observation. 
 
\section{Conclusions}
\label{sec:conclusion}

We have modeled the broadband (1--200~keV), simultaneous \asca-\rxte\ X-ray spectrum taken on September 23--24, 1998, during the hard intermediate state of \j1550, with a model which includes a multitemperature blackbody-emitting accretion disk, a IC-scattering corona, a relativistically smeared reflection spectrum, a more distant, neutral reflection component, and high energy tail. The key result of this modeling is that the inner disk is found to be inclined at an angle $\sim35^{\circ}$ lower than the well-established orbital and jet inclinations $\sim75^{\circ}$ \citep{Orosz2011, Steiner2012}. 

Given the constraints on both the orbital and jet inclinations, and the lack of precession detected in the well-observed transient jet \citep{Corbel2002b}, it is with skepticism that we speculate the inner regions of the accretion disk of \j1550\ could be misaligned with the binary orbit, as a result of spin-orbit misalignment. Recent GRMHD simulations (e.g., \citealt{Liska2018,Liska2019}) do point to a scenario in which the inner regions of the disk may persist to misalign with the outer disk, whereas the jet and corona may torque into alignment much further in. The understanding of how the physics of this process applies to such large misalignments implied by our modeling results is very limited.

 Since there is a lack of supporting evidence for misalignments in the inner regions of the inflow/outflow, we also suggest alternative and arguably more natural explanations: perhaps the inner disk is more vertically extended than the typical thin disk scale heights, given the high luminosity of the hard intermediate state of \j1550, and this leads to geometrical effects which mimic a lower disk inclination. Such geometrical effects may also naturally lead to obscuration of the blue-ward Fe K line emission, also subsuming a lower-inclination disk. Such a geometrical setup would mainly affect the normalization of the disk thermal continuum, and so would yield the same spin measurement already inferred from continuum modeling \citep{Steiner2011}. A further possibility may be that the irradiating source is more vertically, and perhaps also laterally, extended, altering the relativistic effects which cause the complex red and blue wings of the Fe K line. These are all open questions that we will address in future work, since they all require significant improvements to reflection models.

We also measure a high abundance of iron in the accretion disk, and therefore suggest that the disk may have a much higher density than assumed in our modeling, a property of reflection modeling that allows for more solar-like iron abundances. We will explore high-density relativistic reflection models in a forthcoming paper, in which we will show results of a broader modeling campaign on the full set of \rxte\ observations of \j1550, covering a significant portion of the HID presented in Figure~\ref{fig:hid}.  

\begin{figure*}
\includegraphics[width=0.33\linewidth]{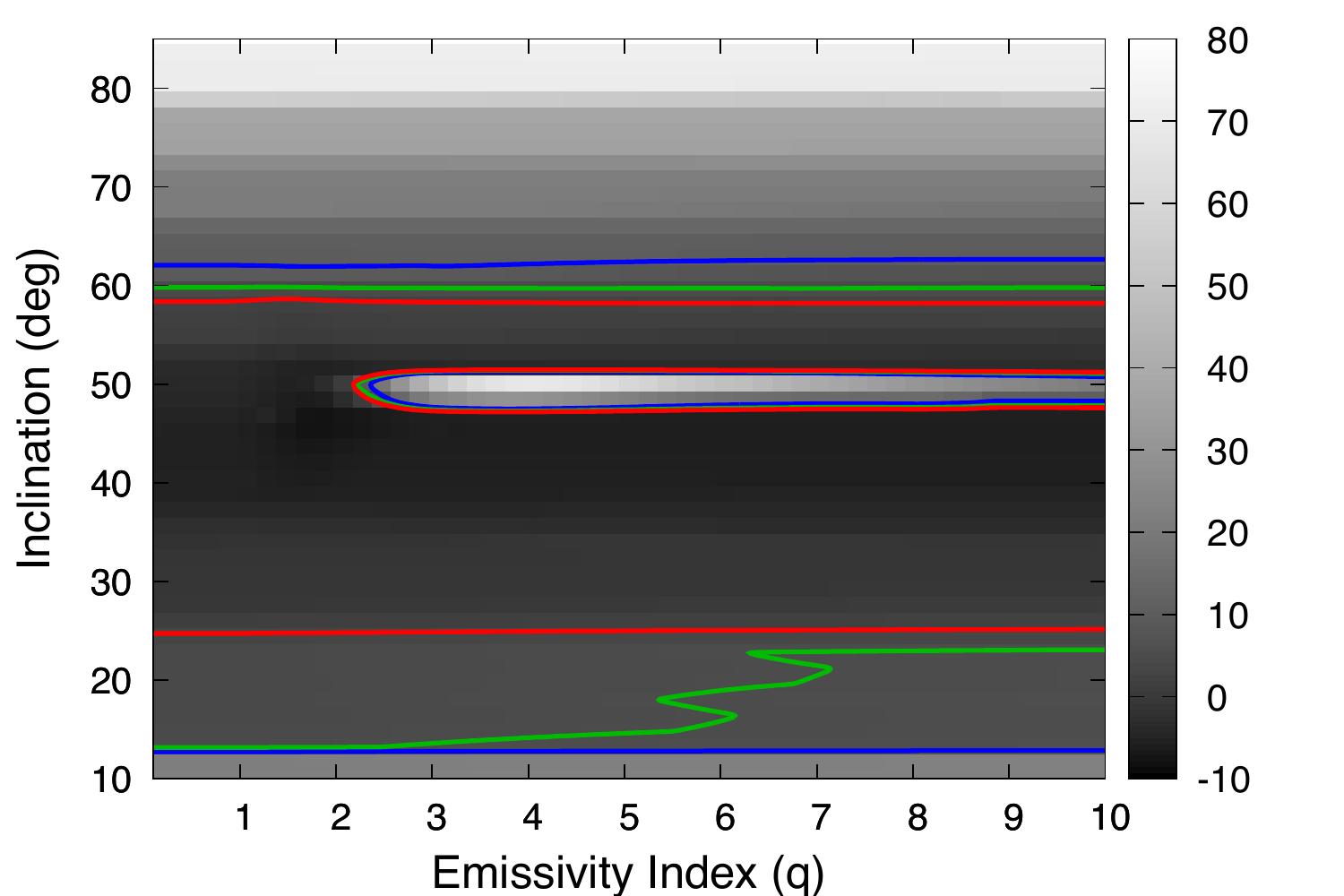}
\includegraphics[width=0.33\linewidth]{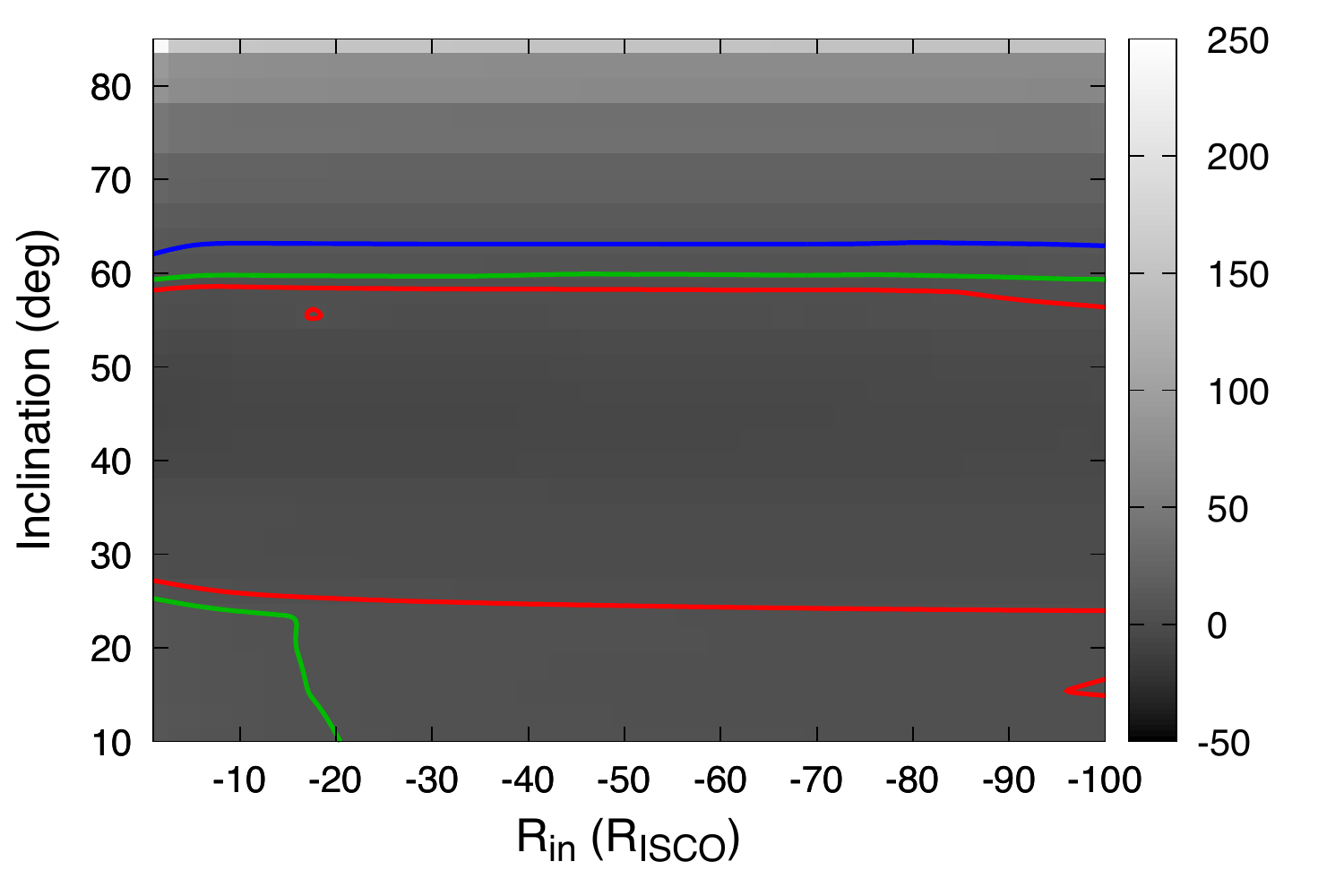}
\includegraphics[width=0.33\linewidth]{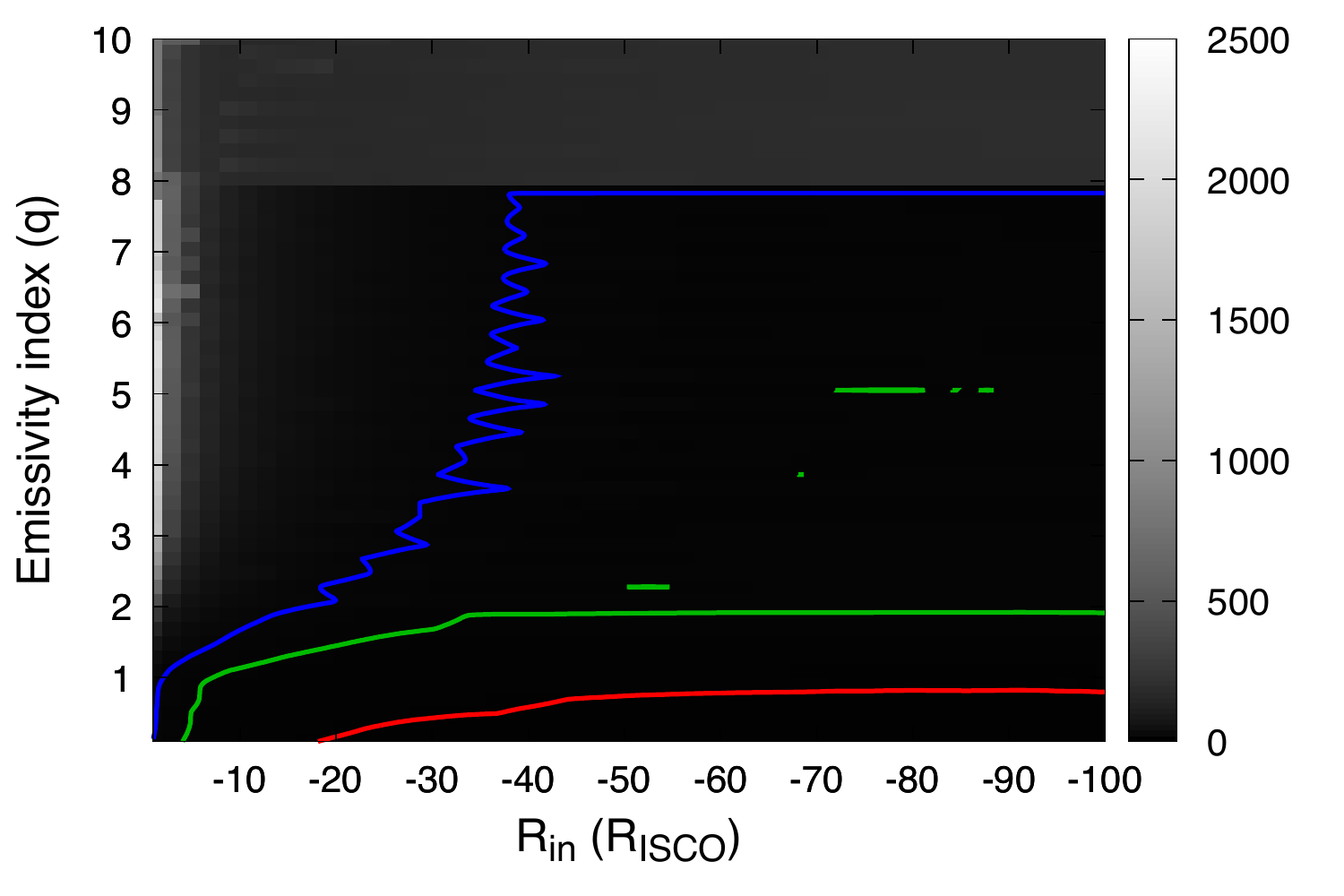}
\caption{The left 2 plots show 2D contour plots of inclination against both the emissivity index ($q$) and inner disk radius ($R_{\rm in}$). Contours shown in red, green, and blue, are the $1\sigma$, $2\sigma$, and $3\sigma$ limits respectively, and the grayscale overlay shows the $\Delta\chi^2$ map. The inclination is constrained to below $\sim60^{\circ}$ at $2\sigma$ confidence. The right hand plot shows 2D contours of $q$ against $R_{\rm in}$ whereby the inclination is fixed at $75^{\circ}$. At $3\sigma$ confidence, both are unconstrained. }
\label{fig:incl_contours}
\end{figure*}

 \acknowledgements
 
 We thank Kristin Madsen, Brian Grefenstette, Hiromasa Miyasaka, Yoshihiro Ueda, and Ken Ebisawa for very valuable discussions regarding the details of the X-ray data used in our analysis. 

 J.A.G. acknowledges support from NASA grant
NNX15AV31G and from the Alexander von Humboldt
Foundation. R.M.T.C. has been supported by NASA
grant 80NSSC177K0515. VG is supported through the Margarete von Wrangell fellowship  by the ESF and the Ministry of Science, Research and the Arts  Baden-W\"urttemberg.
 
This research has made use of data, software and/or web tools obtained from the High Energy Astrophysics Science Archive Research Center (HEASARC), a service of the Astrophysics Science Division at NASA/GSFC and of the Smithsonian Astrophysical Observatory's High Energy Astrophysics Division.

This work has also been conducted within the \nustar\ working group, a project led by the California Institute of Technology, managed by the Jet Propulsion Laboratory, and funded by the National Aeronautics and Space Administration. We thank the support of its members in producing this research. 

\vspace{5mm}
\facilities{\rxte(PCA; \citealt{Jahoda1996}, and HEXTE; \citealt{Rothschild_1998a}), \asca(GIS; \citealt{Makishima1996}), HEASARC}

\software{{\tt XSPEC v.12.10.0c} \citep{Arnaud1996},
		{\tt EMCEE} \citep{fm13}, {\tt XILLVER} \citep{Garcia2010,Garcia2013}, {\tt RELXILL} (v1.2.0; \citealt{Garcia2014,Dauser2014}).}

\appendix
\section{MCMC contours}

\label{sec:app}
Figure~\ref{fig:corner} shows the contours and 1D posterior probability distributions for all parameters in the application of Model~3 to the broadband (1--200~keV) \asca-GIS and \rxte-PCA-and-HEXTE spectrum (Section~\ref{sec:finalfit}), except those associated with the {\tt crabcorr} model. The MCMC routine was initialized with 100 walkers per free parameter uniformly around the best fit values shown in Table~\ref{tab:finalfit}, with normalization parameters distributed with flat log priors. The chain was run for one million steps, and the initial 30\% of the chain was subsequently discarded as the ``burn-in" phase such that the final distributions are fully converged. 

Though there is some skewness to some distributions, most parameter walkers are localized to a Gaussian region encapsulating the best fit value. The strongest notable parameter correlations are between the disk normalization ($\log N_{\rm dbb}$) and disk temperature ($kT_{\rm bb}$), and the ionization ($\log\xi$) and iron abundance ($A_{\rm Fe}$) the disk reflection component. 

\section{Inner radius and emissivity}
\label{sec:app2}

In Section~\ref{sec:finalfit} we describe an exhaustive fit of Model~3, our final spectral model, to the broadband 1--200~keV \asca/\rxte\ spectrum of \j1550. In that fit, the results of which are shown in Table~\ref{tab:finalfit}, we fixed both the disk inner radius ($R_{\rm in}=R_{\rm ISCO}$), and the emissivity index for the illumination of the disk ($q=3$). Here we instead show the results of allowing both parameters to vary freely, in an effort to search for any degeneracy between either $R_{\rm in}$ or $q$ and the disk inclination. Figure~\ref{fig:incl_contours} (left two panels) shows 2D contours between both $q$ and $R_{\rm in}$ and the disk inclination, the results of running a {\tt steppar} $50\times50$ gridded parameter exploration between two variables. The results show that at the $2\sigma$ confidence level, the disk inclination must be less than $\sim60^{\circ}$ across all truncation radii from 1--100 $R_{\rm ISCO}$, and emissivity indices ranging from 0--10. In addition to the inclination limit, both $q$ and $R_{\rm in}$ are unconstrained, such that one cannot distinguish these results from those obtained in Section~\ref{sec:finalfit} whereby both values were kept fixed at $q=3$ and $R_{\rm in}=R_{\rm ISCO}$.

The right hand panel in Figure~\ref{fig:incl_contours} shows contours resulting from a $50\times50$ {\tt steppar} grid between $q$ and $R_{\rm in}$ in which the inclination is fixed at $75^{\circ}$. At $2\sigma$ significance, $R_{\rm in}$ is constrained to beyond a few $R_{\rm ISCO}$, but at the $3\sigma$ level its value, as well as that of $q$, is unconstrained. The best achievable fit to the data with $i=75^{\circ}$ is $\chi^2=1113/877=1.27$.

\bibliographystyle{aasjournal}
\bibliography{references}
%
%
%
%

\end{document}